# A Social Cognitive Heuristic for Adaptive Data Dissemination in Mobile Opportunistic Networks

Matteo Mordacchini[a,∗], Andrea Passarella[a], Marco Conti[a]

[a]*Institute for Informatics and Telematics, National Research Council, Pisa, Italy*

**Abstract**

It is commonly agreed that data (and data-centric services) will be one of the cornerstones of Future Internet systems. In this context, mobile Opportunistic Networks (OppNets) are one of the key paradigms to efficiently support, in a self-organising and decentralised manner, the growth of data generated by localized interactions between users mobile devices, and between them and nearby smart devices such as IoT nodes. In OppNets scenarios, the spontaneus collaboration among mobile devices is exploited to disseminate data toward interested users. However, the limited resources and knowledge available at each node, and the vast amount of data available in the network, make it difficult to devise efficient schemes to accomplish this task. Recent solutions propose to equip each device with data filtering methods derived from human information processing schemes, known as Cognitive Heuristics. They are very effective methods used by human brains to quickly drop useless information and keep only the most relevant information. Altought cognitive-based OppNet solutions proved to be efficient (with limited overheads), they can become less effective when facing dynamic scenarios or situations where nodes cannot fully collaborate with each other, as we show in this paper. One of the reasons is that the solutions proposed so far do not take take into account the social structure of the environment where the nodes are moving in. In order to be more effective, the selection of information performed by each node should take into consideration not only the relevance of content for the local device, but also for other devices will encounter in the future due to mobility. To this end, in this paper we propose a social-based data dissemination scheme, based on a cognitive heuristic, known as the Social Circle Heuristic. This heuristic is an evaluation method that exploits the structure of the social environment to make inferences about the relevance of discovered information. We show how the Social Circle Heuristic, coupled with a cognitive-based community detection scheme, can be exploited to design an effective data dissemination algorithm for OppNets. We provide a detailed analysis of the performance of the proposed solution via simulation.

## 1. Introduction

The evolution of device-to-device (D2D) technologies for communications between mobile devices, and the increasing ability of such devices to collect, store and elaborate relevant amounts of data, is pushing forward the possibility of moving Future Internet functions and services from remote, centralised core network or cloud platform operators to the edge of the network. In this scenario, users' mobile devices will be continuously immersed in a pervasive "data space", where they will have the opportunity to exchange a huge amount of data with other devices and smart "things" nearby. Since this data could be of interest for many users dispersed throughout the network, one of the emerging key challenges is the design of effective,

∗Corresponding author
*Email addresses:* `matteo.mordacchini@iit.cnr.it` (Matteo Mordacchini), `andrea.passarella@iit.cnr.it` (Andrea Passarella), `marco.conti@iit.cnr.it` (Marco Conti)

efficient, scalable and decentralised mechanisms that, through the direct exchange of locally available data between devices, optimise the information dissemination process toward interested users at the entire network level. With respect to these issues, Opportunistic Networks (OppNets) are nowadays one of the most popular paradigms for supporting direct D2D communications in self-organising mobile networks. In Opportunistic Networks there are no precomputed paths from sources to destinations. Rather, physical encounter events between nodes are opportunistically exploited to exchange data. In fact, nodes evaluate how suitable is another encountered node to bring data (closer) to interested users. In this way, opportunistic networks do not face some of the typical problems related to the instability of mobile networks that affect MANET solutions. OppNets are actively investigated by the research community since almost ten years now, and are likely to have a significant impact thanks to the standardisation of D2D Proximity Services (ProSe) in forthcoming LTE and 5G releases[1]. More in general, Opportunistic Networking is considered one of the basis for a number of technologies and applications [1, 2], such as traffic offloading, communication in challenged areas, censorship circumvention, proximity-based applications. With respect to the latter, Opportunistic Networking is one of the elements of crowd sensing solutions, which is one of the most important evolutions of general sensor networks [3] considered in recent years [4].

In this paper, we describe and evaluate a OppNet data dissemination mechanism based on completely self-organising and distributed algorithms running on mobile devices. The contribution of this paper is the design and evaluation of an OppNet data dissemination scheme that allows each node to perform the data selection task by taking into account the needs of the social enviroment in which the user moves. Specifically, we exploit a scheme based on human *cognitive heuristics* [5, 6]. Cognitive heuristics are simple models of the cognitive processes of the human brain derived in the cognitive psychology literature. The main intuition behind the use of cognitive heuristics is that the problem faced by a node in OppNet data dissemination closely resembles what our brain constantly has to do to acquire, retain, drop and spread information coming from the surrounding physical environment. Moreover, according to the cyber-physical convergence view [7], personal mobile devices (such as smatphones, tablets, wearable devices, etc.) can be seen as *proxies* of their human users in the cyber world, since, most frequently, it is through these devices that users access the vast amount of data available in the cyber world. Building self-organising algorithms on model of the human cognitive processes thus means making mobile devices act in the cyber world as they human users would do if facing the same task in the physical world, and thus allowing them to automatically filter information in the same way their human users would do..

We have recently proposed some solutions for disseminating data in OppNets based on cognitive heuristics (e.g., [8, 9, 10, 11]). However, the data dissemination schemes proposed so far do not exploit any knowledge about the *social structure* of the environment where the users move. As such, nodes typically behave in a greedy way, i.e. they drop information that they consider irrelevant for themselves. While this may be appropriate in some cases, taking into consideration the requirements of other nodes frequently encountered (i.e., of the social context of the users) has proven very useful in data dissemination in general (see, e.g. [12, 13, 14]), and it is thus a direction worth exploring also for cognitive-basd data dissemination schemes.

Based on these remarks, we propose a OppNet data dissemination method that exploits the *Social Circle* [15] cognitive heuristic (SCH). In our scheme, whenever two nodes meet, they decide what to replicate locally out of the data available on the other nodes. This is the basic mechanism through which data disseminates in the network and reaches interested nodes. Specifically, we assume that nodes have a limited amount of storage space allocated to help the dissemination process. Therefore, when they encounter other nodes, they have to decide what data to keep and what data to drop, if the amount of data available is larger than the available storage space. To take this decision, in our scheme each node assesses the relevance of data first for itself. In case this knowledge is not sufficient to take a decision (i.e., there are too many items that are all relevant or irrelevant for the individual), the heuristic assesses the relevance for other people in the individual's social context, ranking people according to their perceived social proximity to the indvidual (i.e., to their belonging to the different social circles of the individual). To this end, each node divides its social contacts into different groups on the basis of their social relevance. Starting from the most socially relevant group, the node ranks the data items using their average relevance for the users inside the group. If needed, the node continues to refine the selection of the available data items using the other social groups, one at a time, until the knowledge coming from these groups is enough to take a decision on which data items are worth storing. In this way, nodes store data items based on the preferences of the social groups they belong to, considering inner social circles first.

---

[1]3GPP LTE Release from 15 on, see http://www.3gpp.org/specifications/67-releases

Similar to existing literature on OppNets, we consider social communities as groups of nodes that physically meet with each other frequently. While other types of definitions are possible, which do not necessarily require physical meetings, there is evidence that mobility and physical encounter patterns are very closely related to social structures, and very often frequency of physical interaction is strongly correlated with social proximity (e.g. [16, 17, 18, 19, 20]). Therefore, in the following, social communities are intended as groups of nodes that meet frequently with each other (see Section 3.4 for a more precise definition). As these nodes are very likely to be socially related, it is reasonable to assume that they are willing to help each other acquiring the data items they need.

We compare the performance of the proposed scheme with respect to other state-of-the-art non-social cognitive schemes, proposed in [8][2]. Results confirm our intuition about the advantage of using social information also in cognitive schemes. The algorithm proposed in this paper, based on SCH, is in general both more effective (i.e., it brings data items to interested nodes) and more efficient (i.e., it does so by generating lower network traffic) with respect to non-social cognitive schemes. More precisely, we show that in static conditions, i.e., when content and interests of users in content do not change over time, the two schemes can be configured to achieve the same data delivery efficiency, but the one using social cognitive heuristics generates lower network traffic. Moreover, in dynamic scenarios, where either new content is generated over time or interests of users change (or both), the advantage of the SCH scheme becomes increasingly evident. Specifically, we show that when new content is generated over time, the non-social schemes enter into a congested state, whereby they generate very high network traffic (compared to the SCH scheme), *without* being able to bring content to nodes that are interested.

Summarising, the main contributions of the paper are as follows:

- we describe a data dissemination algorithm for Opportunistic Networks based on social cognitive heuristics for self-organising, P2P data dissemination in mobile opportunsitic networks;

- we show that using social information to drive the dissemination process provides very significant advantages also when cognitive schemes are used;

- we characterise in detail the performance of the proposed scheme, comparing it with reference non-social cognitive schemes; specifically, we evaluate its performance in static and dynamic scenarios, with or without node selfishness, and for a range of parameters charcterising both the behaviour of the proposed algorithm and the features of the overall network environment.

The rest of this paper is organised as follows. Section 2 presents some of the main results in the area of decentralised, self-organising data dissemination solutions in OppNets. Section 3 presents the SCH scheme in detail. The performance evaluation of this scheme is presented from Section 4 to Section 8. In particular, Section 4 presents the experimental parameters and the simuation scenarios used to evaluate the proposed approach. Section 5 shows a comparison of our scheme against non-social cognitive schemes under different patterns according to which users belonging to different communities meet with each other. Section 6 evalautes SCH under a series of dynamic conditions, while Section 7 presents a sensitiveness analysis of the the proposed approach, and Section 8 analyses the behaviour of the system when nodes adopt partially selfish behaviours to save resources. Finally, Section 9 concludes the paper.

---

[2]We do not compare with non-cognitive schemes, as they have been shown to be less efficient in general than cognitive-based schemes [8].

## 2. Related Work

The closest approaches to the one proposed in this paper are some recent schemes (e.g. [8, 21, 22]) that exploit cognitive heuristics to estimate at each node the relevance of the items to be fetched during encounters between pairs of nodes. These schemes are based on the Recognition Heuristic [5]. Essentially, by applying the recognition heuristic, the relevance of data items carried by encountered nodes decreases with the number of times they have been seen on previously encountered nodes (see [8] for details on the recognition heuristic in general, and how it is applied to OppNet data dissemination). In this scheme nodes store only relevant (i.e., non recognised) data items. This approach proved to be as effective in spreading the information as other non-cognitive based approaches, like ContentPlace [12], which are considered references in the OppNet literature. On the other hand, it is more efficient, as it generates far lower network traffic to achieve the same level of data dissemination [8]. As discussed in Section 1, the main difference with the proposed scheme is that here nodes take into consideration relevance of data items also for the other nodes in their social context, which proves to provide significant performance benefits and applicability to more general scenarios.

A number of solutions for data dissemination in opportunistic networks have been proposed, which do not exploit cognitive heuristics [23]. Note that in [8] we have shown that non-social cognitive heuristics outperform ContentPlace [12], which is one of the most efficient non-cognitive heuristics proposed so far. Therefore, in this paper we do not compare against this class of solutions, as comparison is implicit in the comparison with [8, 21]. Hereafter we provide a quick overview of the main non-cognitive approaches, along the taxonomy provided in [24]. We redirect the reader to that paper for a more complete survey.

The first work about content dissemination in OppNets was developed in the PodNet Project [25]. Data items in the PodNet system belong to general topics, i.e. *channels*. Nodes subscribe to channels, trying to retrieve all its related items. Moreover, nodes devote part of their caches in order to collaborate in the data dissemination process. In fact, upon encounters, in PodNet nodes make use of specific policies o decide which data items to exchange, in addition to those of specific interest for the local users. The authors propose four different strategies, which are functions of the nodes' interests in the various channels. With all the proposed strategies, the performance of PodNet shows that the system is able to outperform a scheme where nodes keep only the items of the channel they are subscribed to.

ContentPlace [12] is an example of social-based data dissemination algorithms. The data dissemination problem is addressed as a multi-constrained knapsack problem. The goal is to maximize the social utility of fetching an item, while taking into account the limited resources of a device. Specifically, in ContentPlace each node estimates the utility of the data items it could potentially fetch from other encountered devices. Then, it ranks the data items it currently stores and those it could possibly fetch according to their utility vales, and selects the most useful ones that fill the amount of local storage space allocated to the data dissemination process. Different definitions of utility have been proposed and evaluated in [12]. Among them, the most useful utility function is a linear combination of a local utility (i.e., the utility of the data items for the local node), and the utility for nodes in social communities the local node will "visit" in the future. Even though ContentPlace is less efficient than recognition-based cognitive heuristic schemes (as shown in [8]) the idea of using information about the relevance of data for *all* the communities to which the local user belongs is also at the basis of the proposed SCH cognitive scheme.

The scheme in [14] is an example of pub/sub scheme also using social information. The main idea proposed in [14] is to identify social communities between nodes. Then, for each social community a broker is determined. Brokers are the nodes with the highest number of social links inside their social community. Brokers form an overlay network, and gossip content availability and content requests to each other, allowing the data items to flow from one community to another, towards the nodes that request them. Non-broker nodes send to the broker(s) of their social communities content requests, which are circulated among all the brokers. Nodes storing data items matching the request are informed by their brokers, and the corresponding data items is sent back (again through the network of brokers) to the requesting node(s).

Another work [26] proposes a dissemination solution for OppNets based on the pub/sub paradigm. The authors' aim is twofold. On the one hand, they proposed an incentive mechanism to increase the participation in the dissemination process of selfish nodes. On the other hand, they propose a solution that evaluate the utility of the data to exchange between nodes upon meeting. Specifically, the utility of a data item is correlated to its freshness. In this way, the system maximises the dissemination of more recent data towards its subscribers.

The scheme presented in [27] exploits an interest-based dissemination solution for OppNets that takes into account the fact that users with similar interests tend to group together. This mechanism is coupled with other informations about the devices, like their contact history and the connections of their users in on-line social networks. These informations are exploited by the authors to derive a series of ranking functions that allow nodes to decide which data to exchange upon contact.

In [28] the authors consider the data dissemination problem as a global optimisation problem. To this end, they propose to view all the nodes' shared memories as a unique, global cache. The paper defines a global utility function, which estimates the best possible allocation of items into all the nodes, considering the items' utilities for each single node, and the expected rate of requests for every item. Each device uses simple local approximation strategies to compute the utility, as global knowledge is clearly not available. Another approach [29] presents a solution that is based on global view of the system. In particular, the authors show how to build a data dissemination multicast tree for OppNets that is specifically optimised to respect delay constraints and communication costs. Since the general solution implies a centralised view of the whole system, the authors also propose a distributed heuristic that allows the nodes to evaluate the suitability of other encountered devices to opportunistically deliver data.

In [30], the authors present a data dissemination scheme for OppNets specifically design for delivering time-sensitive data toward interested users. The proposed approach makes use of three main components to let nodes decide whether to exchange data upon meeting. The system exploit a node's history of visited areas to compute its probability to appear in a specific location at a given time. This component is coupled with the others depinding ont he role of the nodes in the data forwarding process. In fact, data source nodes employ also a relay node selection mechanism to decide the initial nodes that are likely to timely deliver the data. Data carrier exploit also a technique for evaluating whether other encountered nodes are more suitable than them to deliver the data they hold on time. Experimental results show the ability of the system to deliver time-sensitive data efficiently.

The work presented in this paper is an extension of the one proposed in [31]. In particular, we extend [31] primarily through a much more extensive and comprehensive evaluation of the proposed solution. Specifically, we added new evaluations about the sensitiveness of the approach to the dimension of the Opportunistic Cache and the initial distribution of data items among groups of nodes. Moreover, we investigated the behaviour of the system when data items have associated Time-to-Live restrictions (see Sec. 7). In addition, we also present new results about the performance of the proposed approach when nodes do not fully collaborate in the dissemination process. In these experiments, nodes adopt a partially selfish behaviour and limit their engagement in the dissemination process. We investigate two partially selfish schemes: one in which a node does not use the strength of social relationships between nodes to discriminate the other nodes with which data items are exchanged; and one that limits the collaboration on the basis of the social connections between a nodes and its contacts (see Sec. 8). These additional results make it possible to reach, , with respect to [31] , a much deeper understanding of the behaviour of the proposed approach under different conditions and various settings of the main parameters.

## 3. Data Dissemination in OppNets using Social Cognitive Heuristics

### 3.1. Problem Statement and System Assumptions

In the following, we consider a very common scenario for data dissemination in OppNets (previously used, e.g., in [25, 12]). We consider an Opportunistc Network where $N$ mobile devices generate data items, each belonging to a specific high-level topic, termed as *channel*. Each device owner is interested in retrieving all the items that belong to one of the available channels (the one it is *subscribed* to). Contacts between nodes are the *only* way to disseminate data. Therefore, each device shares the items generated locally and reserves a little amount (with respect to the total amount of data items to be disseminated) of its storage space to help the dissemination process. In the following, the storage space that contains the items generated locally is called the Local Items cache (LI), while the storage space used to collaborate in the dissemination process is called the Opportunistic Cache (OC). Received data items of the subscribed channels may be kept in the OC based on the decisions of the SCH scheme presented hereafter. Otherwise, they are supposed to be dropped after being consumed by the local user.

The core of the data dissemination scheme is therefore the policy a node uses to select items to keep in the OC upon encountering another node. Before presenting this policy, in the following section we describe the social circle heuristic upon which the policy is based.

*3.2. The Social Circle Heuristic*

It has been shown that the human brain exploits not only the individual's judgement about relevance of information, but also estimates the judgement of social peers to derive accurate decisions about which information to keep [15]. In the cognitive sciences, the *Social Circle Heuristic* (SCH) [15, 32] is proposed as a model for the cognitive mechanisms used by the brain to this end.

As many of the other cognitive heuristics (see [8] for a brief summary), SCH is essentially a decision-making scheme. It is defined as an algorithm used by the brain to take a decision about a given choice (for illustration purposes, in the following, we will refer to whether to keep an information in memory as an example, but SCH is clearly applicable to many different decision problems). As all the other cognitive heuristics, SCH limits the information and the computation (i.e., the use of cognitive resources) needed to take this decision. To this end, it uses a series of decision steps in a specific order. The process terminates at the first step that is sufficient to discriminate between the possible choices.

The first step used by SCH is applying another cognitive heuristic, i.e. the Recognition Heuristic [33]. In general, the Recognition Heuristic (RH) discriminates between two pieces of information $A$ and $B$, based on whether one of them is recognised and the other is not. An information is recognised if it has been seen in the environment more than a fixed number of times (a recognition threshold). RH corresponds to using only local information to assess relevance, and is indeed the cognitive heuristic used in previous work in the literature [8, 21]. The following steps taken by SCH are those where it considers social information into account. They are used if the RH step is not sufficient to discriminate between the possible choices, and therefore to take a decision (in our example, which information to keep between $A$ and $B$, because both are either recognised or not recognised).

In the "social" steps, SCH exploits the fact that each person's social contacts can be grouped in different clusters (or circles), according to the strength of their relationship with the individual [34, 35], using the individual itself as the first social circle. Assuming that such circles can be identified (we will come back to this point in Section 3.4), SCH considers sequentially each circle in order of social strength. In general, at each step, it assesses the possible choices for the members of the considered social circle. As soon as considering one circle results in a sufficient discrimination between the possible options (e.g., one among $A$ or $B$ is considered more important for the members of a social circle), SCH stops. To assess the possible choices of the members of a circle, typically RH is used also in these steps. It is applied for all nodes in the social circle, and the individual then aggregates the results, for example by considering an information recognised if a majority of members of the social circle have recognised it.

It is worth noting that SCH analyses the circles' preferences in order of social closeness, and often terminates before looking through all the social circles. As a consequence, the influence of more peripheral social circles is typically much weaker than the influence of most proximal social circles.

*3.3. SCH for Data Dissemination in OppNets*

In this section, we show how SCH can be applied by nodes in an OppNet to select relevant information upon encountering other nodes. As the algorithm involves several steps and operations, we first provide a high level overview of the proposed approach in Section 3.3.1, while in Section 3.3.2 we introduce the main concepts behind the cogntive Recognition Heuristic, i.e. one of the main building blocks of our solution. The detailed description of the approach is given in Section 3.3.3.

*3.3.1. High Level Overview*

When two nodes encounter, they exchange summaries of the items they hold in their data caches, i.e. the LI and OC caches. Without loss of generality, let us focus on one of the two nodes, and explain the algorithm from this node's perspective.

At this stage, a node has to decide which data items it has to carry in its OC among the ones it already stores and the ones it could fetch from the encountered device. The node performs this task by evaluating the relevance of the available data items using SCH. Each step of SCH allows the node to define a so-called *consideration set*, which represents the relevance of data items under scrutiny for one of the social communities the node belongs to. Specifically, as it will be explained in the following sections, SCH analyses social communities in decreasing order of social proximity to the node. The concept of consideration sets comes from the the cognitive psychology literature, where it indicates intermediate results in multi-alternative choices [36]. In the application of SCH, the data selection process is ended in case a consideration set obtained through a step of SCH fits the size of the node's OC. Whenever the size of the consideration set is larger or smaller than the OC size, further SCH steps are performed in order to prune or add data, respectively. At the end of the process, the node fetches from the other node the data items in the final consideration set that it does not hold locally.

*3.3.2. The Recognition Heuristic*

Since the evaluation of data items through RH is one of the building blocks of all the steps of SCH, we give a separate description of how RH can be implemented in OppNets. The following description is in accordance with the method we first proposed in [8].

In order to evaluate a set of data items, a node takes into account the items themselves along with the channels they belong. Precisely, a node considers that a data item is relevant **if and only if** its channel is recognised and the item itself is not recognised (we explain the rationale of this choice in a moment). In order to recognise channels and items, we use the precise recognition algorithm defined in the cognitive literature [5]. A node maintains separate counters for each channel and for each data item it is aware of. A channel counter keeps track of the number of *distinct* subscribers of the channel that have been met by the node. An item counter tracks the number of times a given item was seen in the caches of other encountered nodes. Channels and items are considered "recognised" in case their associated counters reach specific *recognition thresholds*. Note that the higher a channel counter, the higher the number of its subscribers. On the other hand, the higher an item counter, the higher the number of copies of that item already spread in the system. This is why an item is considered relevant if its channel *is* recognised, and if the item itself is *not* recognised. Hereafter, the value of the counters are also called *recognition levels*.

*3.3.3. Detailed Description of the SCH Algorithm*

The precise SCH-based data exchange process when two nodes encounter is described in detail in Algorithms 1 and 2. Without loss of generality, we explain the operations of the algorithm from the standpoint of one of the two nodes. We begin the presentation of the process from Algorithm 1, executed by the node when the contact with the other node starts. This algorithm starts by considering the items available on the local OC together with the items in the encountered node's LI and OC (lines 1-3). This makes up a set $P$, which is the starting point for the selection process. The first selection performed on this set by SCH exploits RH, using the method described in Sec. 3.3.2 (lines 4-11). To this end, in Algorithm 1, function $r_{lev}$ (line 8) gives the recognition level of an item $i$ ($r_{lev}(i)$), or of its channel ($r_{lev}(i.ch)$), while conditions in line 8 implement the recognition policy described in Sec. 3.3.2. In case RH does not suffice, the SCH social steps are exploited (lines 12-16). After the first step, the cardinality of the consideration set created so far could exceed the dimension of the OC, or it can be smaller than that size. We first present the behaviour of the system when the size of the consideration set is larger than that of the OC (lines 12–13), while we discuss the case when the size of the consideration set is smaller than the size of OC at the end of this section (lines 14–15). In both cases, the same procedure can be used. This is represented in Algorithm 1 by the function $sch()$, which is called with different parameters in the two cases. The definition of function $sch()$ is presented in Algorithm 2.

When the size of the first consideration set is larger that the OC size, the system performs the social steps of SCH. To this end, we assume that each node divides its social contacts into distinct groups exploiting the cognitive-based community detection scheme defined in [37]. We give a brief description of this algorithm in Sec. 3.4. As a result of the community detection algorithm, each node maintains a set $G = \{G_0, \ldots, G_n\}$ of social groups, where $G_0$ is the node itself, and $G_1, \ldots, G_n$ are ordered on the basis of social importance of those users for the node. More precisely, users are grouped on the basis of the *strength of their social ties* with the node (line 1 of Algorithm 2). Algorithm 2 is in charge of pruning the items in a consideration set, by recursively using the information of the social circles in $G$. Specifically, Algorithm 2 receives as input i) a set of data items to be pruned; ii) the remaining space to be filled in the node's OC; iii) the social group to be used to evaluate the data items. The relevance of each data item is assessed using the same general scheme of RH shown in Section 3.3.2. Thus, data items are relevant if their channels are recognised and the data items themselves are not recognised. However, in this case the algorithm does not use the channels and items recognition levels of a single node to take decisions. Rather, it exploits the *mean* recognition levels of both channels and items inside a given social group. As a consequence, a channel (equivalently, a data item) is considered as recognised if the *mean value* of its recognition level among the nodes in the social group is above the recognition threshold. A node is able to compute the mean recognition levels for its social groups since we assume that, upon contacts, nodes exchange their individual recognition levels for each channel and each data item, along with their own data summaries. In Algorithm 2, $S$ is the consideration set to be filtered, $O$ is the space still available in the OC and $j$ is the index of the social group used to filter the data items. Groups with lower indices are the ones with stronger connections with node that is carrying out the evaluation. The filtered data items are returned in the set $F$ (line 16). The selection based on the recognition level conditions is done in lines 8–9 of the pseudo-code. Items are divided into a series of sub-groups $S_1, \ldots, S_k$, where all the items in a sub-group $S_i$ have the same mean item recognition level (line 8) and, consequently, they all have the same relevance inside the social group $j$. Data items are placed in the final consideraton set $F$ starting from those with the lowest average item recognition level (i.e. $S_1$) and contuining with the others (lines 9–13), until adding all the data items of a group $S_i$ would lead the size of $F$ to exceed the size limit $O$ (line 10). Some of the items in $S_i$ could be selected. However, they are all equally relevant in social group $j$. Thus, they are further evaluated using the knowledge of other social groups. To perform this evaluation, Algorithm 2 calls recursively itself (line 11). In this recursive call, the parameters are set in the following way. The consideration set to be pruned becomes $S_i$, since data items with lower recognition level are already in the final consideration set and will be kept in OC, while items with higher mean recognition levels should not be considered, since even limiting to $S_i$ already overflows the OC. The group of nodes used for further assessment is the next most closely related social group, i.e. the one with index $j + 1$. The space that could be filled is the difference between the one originally available and the dimension of the consideration set selected so far, i.e. $O - |F|$. Finally, in case the knowledge coming from all the node's social communities is not enough to filter the data, lines 4–6 are used as a last resort choice, in accordance with the cognitive psychology literature about SCH [15]. In fact, a uniform random selection is done on the final set of items to be discriminated.

Finally, a special case for Algorithm 1 happens when the total number of items filtered by the first step of SCH is lower than the capacity of the OC. Since the OC is thought to be a space devoted to help the overall data dissemination process, we allow the node to possibly revise its individual judgement about the utility of discarded items (line 14–15 of Algorithm 1). Specifically, the node considers its social circles, looking whether items that it would discard (based on the first step of SCH) can be considered of sufficient social relevance to be kept in OC. In this case the individual node is not considered, and the selection process starts from the social community $G_1$ (line 15).

### 3.4. Cognitive-based Community Detection

In order to exploit the Social Circle Heuristic, nodes detect their social communities using the cognitive-based algorithm presented in [37]. This solution exploits the cognitive notion of memory activation. For a given node, the activation level of another node is a function (defined in [38]) of frequency and recency of contacts with that node. Intuitively, activation is higher if contacts are more frequent and more recent. As shown in [38], to compute activation values, each node only needs to monitor the inter-contact times with other nodes, i.e., the time elapsed between consecutive contacts with the same node. Activation is a simple function of the inter-contact times. In [37], each node uses activation values of other nodes to cluster them into social communities. Clusters are also computed using cognitive algorithms taken from [39]. Results presented in [37] show that the this community detection algorithm is effective, is able to track the dynamic evolution of physical encounters into a corresponding dynamic update of social communities membership, and requires minimal information to be exchanged between nodes (thus resulting in a very limited network overhead).

**Algorithm 1** Social-based Data selection exec. by a node $n$
---
1: Let $I$ be the set of items owned by an encountered node
2: Let $O$ be the max space available in the OC
3: Let $P = I \cup OC$ be the set of items to be pruned
4: Let $\theta_C$ be the channel recognition threshold
5: Let $\theta_I$ be the item recognition threshold
6: Let $S = \emptyset$
7: **for** all $i \in P$ **do**
8:    **if** $r_{lev}(i.ch) \geq \theta_C$ **and** $r_{lev}(i) < \theta_I$ **then**
9:       $S = S \cup \{i\}$
10:    **end if**
11: **end for**
12: **if** $|S| > O$ **then**
13:    $OC = sch(S, O, 0)$
14: **else if** $|S| < O$ **then**
15:    $OC = sch(P - S, O - |S|, 1)$
16: **end if**

**Algorithm 2** Function sch(S,O,j)
---
1: Let $G = \{G_0, \ldots, G_n\}$ be the set of social groups
2: Let $S$ be the set of item to be filtered
3: Let $O$ be the max space available left in the $OC$
4: **if** $j > |G|$ **then**
5:    **return** $O$ randomly chosen items from $S$
6: **end if**
7: Let $F = \emptyset$
8: Let $S' = \bigcup_{k=0}^{\theta_I} S_k$, where $S_i = \{s \in S | r_{lev}^j(s) = i \bigwedge r_{lev}^j(s.ch) > \theta_C\}$
9: **for** all $S_i \in S'$ s.t. $i < \theta_I$ **do**
10:    **if** $|F \cup S_i| > O$ **then**
11:       **return** $F \cup sch(S_i, O - |F|, j+1)$
12:    **else**
13:       $F = F \cup S_i$
14:    **end if**
15: **end for**
16: **return** $F$

## 4. Performance Evaluation Scenarios

In the following sections, we report a series of results about the algorithm performance obtained by simulation under various scenarios. To do that, in this section we present the main components that are used to create the simulation and evaluation environments. In particular, in Section 4.1 we introduce the mobility model used to simulate the movement of nodes in the simulation area. In Section 4.2, we present the scenarios that are used to determine the social grouping and the connections between nodes, along with and the main parameters that are exploited to evaluate the proposed solution. Finally, Section 4.3 presents the performance indices and the strategy used for the evaluation of the system.

*4.1. Nodes Mobility Model*

Node mobility is simulated using the HCMM model [40]. HCCM is recognized by the OppNet research community as one of the reference models for simulating the mobility of nodes in a Mobile Network [41, 42, 43]. In particular, one of the salient features of HCMM is its ability to closely approximate the human mobility using the social relationships existing between the users in the environment. In fact, HCMM integrates temporal, social and spatial notions in order to obtain a realistic representation of real users movements. In order to achieve this goal, its design is inspired by results in the sociology and complex networks literature. One of its main features is the ability to reproduce statistical properties of real user movement patterns, such as inter-contact times and contact durations. The focus of HCMM on reproducing human contact and inter-contact times is particularly important in opportunistic networking. Other mobility models in the literature are either not so realistic from this standpoint [44], or are more amenable to different types of networks with repect to OppNets, such as, e.g., cogntive radio networks [45].

In HCMM, the simulation space is divided in cells. Each cell could host a group of nodes, that represent a social community of users that are physically co-located. Each social group is initially assigned to a cell (its *home-cell*) avoiding that two groups are physically adjacent (no edge contacts between groups) or in the same cell, to avoid physical shortcuts between groups. Nodes move inside their home cell, according to a random process. Specifically, a node selects a destination inside the home cell according to a uniform distribution over all possible points inside the cell. The speed is also selected according to a uniform distribution over the speed range. Some special nodes exist, called *travellers*, which not only move inside their home cell, but can also "visit" other cells, thus bridging between distinct communities. Specifically, a traveller selects the community to "visit" based on social relationships with members of each community (other than its home one). When the traveller moves outside its home cell, it uses the same algorithm described above to select the point inside the cell where to move to. After each movement, it decides, according to a Bernoulli random process with a given probability $p$, whether to stay visiting the community longer, or going back to its home cell.

*4.2. Simulation Scenarios*

As stated in the previous section, travellers are the only mean of connection between communities. Therefore, they play a crucial role in the data dissemination process, since data diffusion in distinct groups can occur only due to travellers bringing messages from community to community. Thus, in order to thoroughly analyse the system performance, in the first part of the evaluation, we test the proposed approach under three different scenarios of social connection between communities. Precisely, in the first scenario we consider a ring topology. In this scenario, each community has only one traveller, that connects it only to the next community: the first community has a traveller to the second, the second has a traveller to the third, and so on (the last has a traveller to the first). Hereafter, we refer to this scenario with OT (i.e. One Traveller per community).

In the second scenario, each community has a traveller that is able to visit all the other communities. However, every time it exits its home community, it selects the destination community using a Zipf probability distribution. As a result, each traveller visits more frequently one community, gives less preference to a second one, and goes rarely to the third one. Also in this case, traveller preferences are rotated across communities to ensure a globally uniform visiting pattern. In the following, this scenario is termed as ZT (Zipf travellers).

In the last scenario, each community has three travellers, each connecting it to one of the other communities. This scenario is called TT (i.e. Three Travellers per community).

The OT and TT scenarios are rather extreme cases. In OT a community has just one outgoing traveller and lacks a direct connection with some of the other groups. As a consequence, all the communities have to heavily rely on travellers belonging to other groups and the information dissemination process is made more difficult. On the other hand, in the TT scenario each community has its own traveller toward all the other communities. We expect a easier (and faster) information dissemination process in this case. ZT represents a more common scenario, where each traveller has a skewed visiting preference about the groups outside its home community. This behaviour better reflects real social contact probabilities in human social relationships [16]. Note that, althuogh limited to 4 social communities, our evaluation settings are representative of a significant range of socially-driven movement patterns, and allow us to thoroughly assess the performance of social circle heuristics.

Table 1: Main simulation parameters

| Simulation Parameters | |
|---|---|
| Simul. Area | 1000m x 1000m |
| Grid | 4x4 |
| Num. of Channels | 4 |
| Items per Channel | 100 |
| Numb. of Communities | 4 |
| Numb. of Nodes | 100 (25 per comm.) |
| Node speed | unif. in $[1, 1.86]m/s$ |
| Transm. range | 20m |
| Simulation time | **50,000s** and 100,000s |
| OC size | **10**, 5, 2 slots |
| Channel Rec. Thr. | 3 |
| OT scenario | chain topology |
| ZT scenario | **Zipf visiting pattern** |
| TT scenario | all communities connected |

Moreover, in all the following experiments, channels subscription popularities within each community are skewed and follow a Zipf distribution with parameter 1. Popularities are rotated among groups. Thus, the first channel is the most popular within group 1, while channel two is the most popular in group 2, and so on. As a result, all the channels have the same total number of subscribers.

Table 1 shows the values of the main simulation parameters used for the simulations for all the scenarios. When more than one value is used in the simulations, "standard" values are put in bold face.

*4.3. Performance Indices and Evaluation Strategy*

In all the experiments reported in the following sections, we evaluate the performance of the proposed solution (SCH in the figures) against a pure recognition-based (RH) cognitive data dissemination system for OppNets [8]. As performance figures we use the Hit Rate and the overhead.

At a specific point $t$ in time, the Hit Rate is defined in the following way:

$$HR(t) = \frac{1}{N} \sum_{n=1}^{N} \frac{n.SC(t)}{n.c} \qquad (1)$$

Thus, the Hit Rate is the mean ratio over all the nodes between $n.SC(t)$, i.e. the number of the subscribed channel's items retrieved so far by a node $n$, and $n.c$, i.e. the total number of objects in the node's subscirbed channel.

The overhead is the total number of messages exchanged in the whole network (up to that point in time), including both data items and control messages (to implement the cognitive heuristics)[3]. All the reported

---

[3] On purpose we do not report the overhead in terms of bytes, because this would depend a lot on the sizes of the different

results are the mean of 10 different runs, obtained using 10 different mobility traces generated with the HCMM model.

In Section 5, we first evaluate the performance under static conditions, i.e. when data items and interest of nodes do not change over time. Then, from Section 6 to Section 8, we give a thorough analysis under the more realist social scenario, i.e. the ZT scenario. In particular, in Section 6, we evaluate the performance in various dynamic conditions. We anticipate that, while SCH always outperforms RH, its performance gains are even greater in dynamic scenarios (which are, clearly, even more realistic). In Section 7.1 we analyse the impact of the OC cache size on the dissemination process, while in Section 7.2 we evaluate how the initial distrbution of data items among groups could affect the performance of the system. Section 7.3 reports the results obtained when data items have a limited TTL. Finally, Section 8 presents the evaluation of the system when nodes have partially selfish behaviours, such that they do not exploit all of their encounters to exchange data and information.

## 5. The impact of different social connection patterns between communities in static configurations

In this section, we compare the performance of SCH and RH in cases of static configurations, as a function of the connectivity patterns between communities.

Figures 1(a)–3(b) report the performance of the two approaches under the OT, ZT, and TT scenarios, respectively. The OT scenario (Figures 1(a) and 1(b)) is the most difficult one for the information dissemination process, that proceeds more slowly with respect to the other cases (i.e., the HitRate curve increases more slowly over time). Two configurations of SCH and RH are considered, corresponding to different values of the respective recognition thresholds (RT in the figure). Let us consider the configurations with higher RT first (12 for SCH and 75 for RH). These values are selected to make SCH and RH stabilise approximately at the same Hit Rate. As shown in Figure 1(b), this is paid by RH with a higher overhead, which is a side effect of the higher value of the item recognition threshold (RT) needed to reach that HitRate. A higher RT value means that in RH nodes keep considering data items relevant for a longer time, and continue exchanging them. On the other hand, in SCH items are recognised faster at each step of the heuristic, and keeps exchanging them only if they are considered relevant for some of the node's social communities. Thus, the majority of the nodes (the non-traveller nodes) stop exchanging data items relatively soon (with respect to what happens in RH) after they have appeared in their social community, while travellers keep bringing data items across communities. On the other hand, SCH and RH can be configured to obtain (approximately) the same overhead (corresponding to RT=3 for SCH and RT=30 for RH). This, however, results in lower HitRate for RH. We anticipate that this behaviour with respect to the RT value constantly appears in all tested scenarios. In general, the OT scenario is the one where the performance gain of SCH over RH is lower, though still evident.

Figures 2(a)–2(b) show the performance of the two solutions under the ZT scenario. In this case, SCH with RT=3 achieves a HitRate of 100%. To get the same performance with RH, we need to increase the RT value up to 75 (and it can be noticed that 100% HitRate is achieved even slightly later). However, this is paid with a much higher overhead. Reducing the RT value to 30 results in a small though noticeable performance loss in terms of HitRate (of about 5%). However, note that the overhead of RH still remains higher than the SCH overhead. With the same value of RT used by SCH, RH obtains a much lower overhead, but the performance drop in terms of HitRate is huge.

The results for the TT scenario are presented in Figures 3(a)–3(b). Results are quite similar to the ZT scenario, with the notable difference that now, as the communities are more connected, RH is able to reach 100% HitRate also with RT=30 (HitRate curves for SCH and RH with RT=30 and 75 overlap). Again,

---

messages. In general, for reasonable values of the size of control traffic (e.g., 4B for items ID and recognition values, 4B for bitmaps of recognised items and channels) the difference between RH and SCH overheads *is further amplified* - with respect to the difference in terms of number of messages - when the average data item size is above 2MB, which is very common for OppNets and DTN traffic in general.

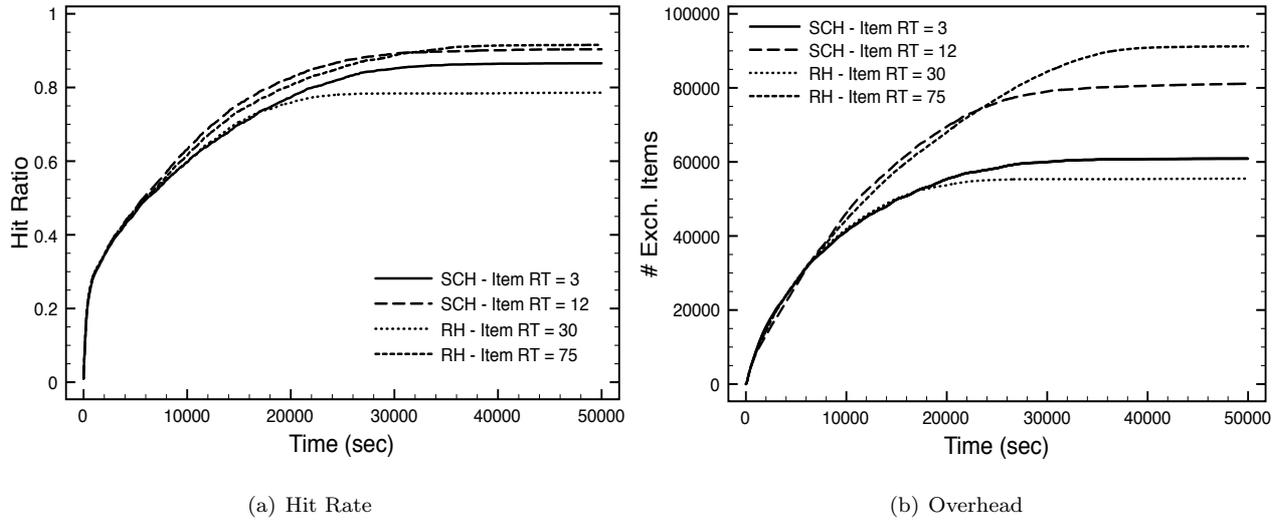

(a) Hit Rate

(b) Overhead

Figure 1: OT Scenario - Hit Rate (a) and Overhead (b)

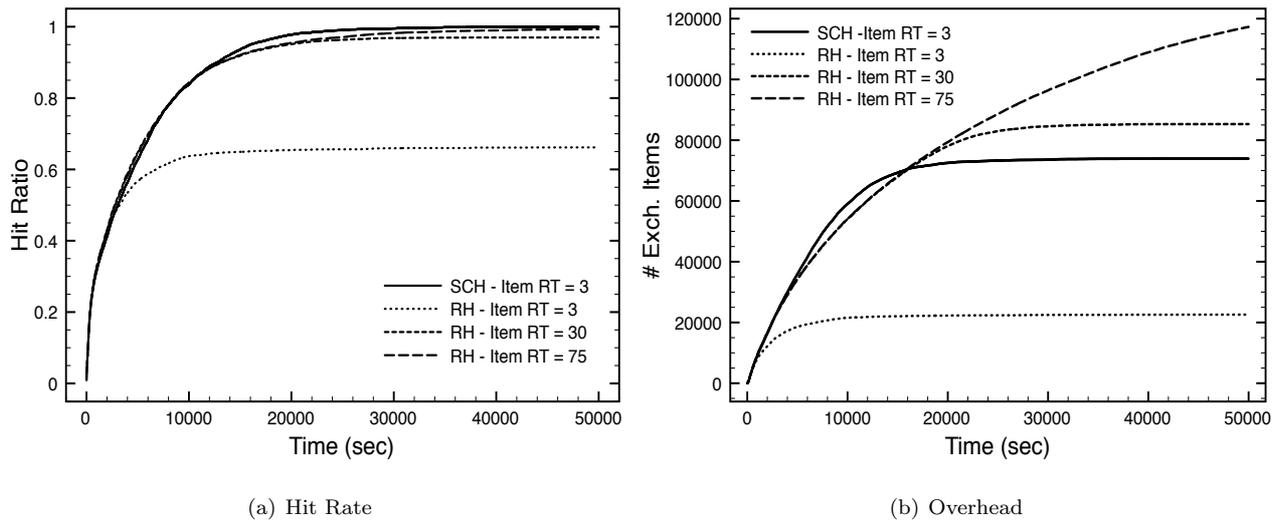

(a) Hit Rate

(b) Overhead

Figure 2: ZT Scenario - Hit Rate (a) and Overhead (b)

SCH achieves the same HitRate with a much lower RT (equal to 3), which results in lower overhead. Using the same RT in RH reduces the overhead significantly, at the cost of far lower HitRate.

The overall conclusion from this set of results is that, in general, to achieve the same hit rate of SCH, the corresponding configuration of RH generates a higher overhead, and this holds for a number of different nodes mobility patterns. We anticipate that, while this behaviour appears also in these statics configuration, it is much more evident in more dynamic environments, which we consider in the next sections.

## 6. Dynamic content generation

We now consider more dynamic conditions. In particular, we investigate the system performance in three different dynamic scenarios. In the first one, nodes abruptly change their subscription. In the second scenario, a new channel (and related items) is suddenly introduced and some nodes subscribe to it. In the third scenario, new data items are generated for each channel. We report the results of this investigation only under the ZT scenario, which is more broadly representative, as discussed before. Qualitatively similar results are obtained also in the OT and TT scenarios.

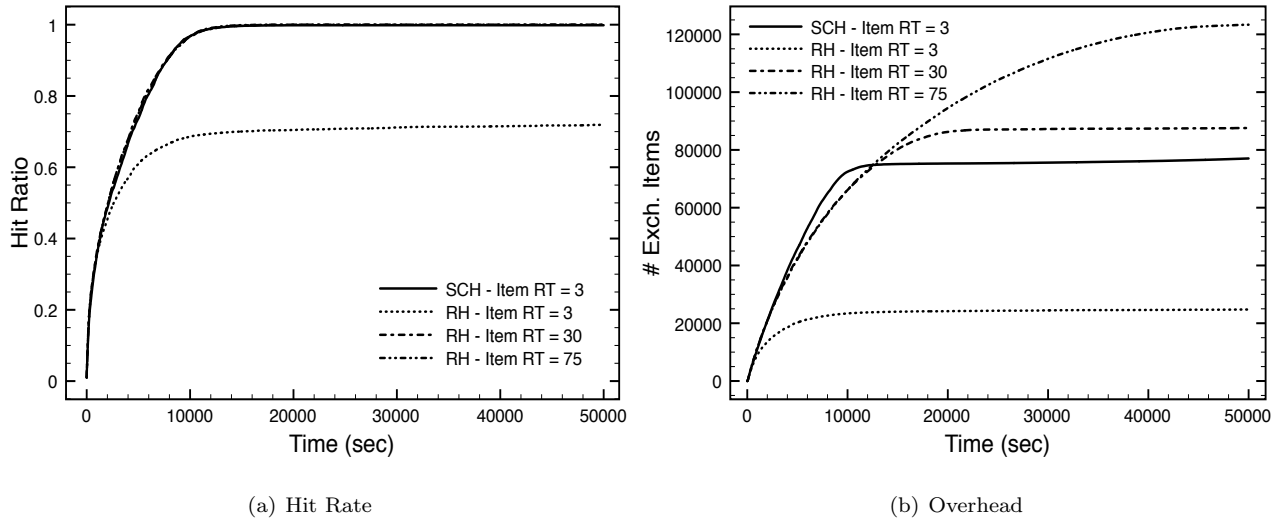

(a) Hit Rate

(b) Overhead

Figure 3: TT Scenario - Hit Rate (a) and Overhead (b)

Figure 4 shows the comparison between SCH and RH in a situation where after some time (10,000s in this case), nodes inside each community (with the exception of travellers) change their subscriptions. Nodes subscribed to channel 1 change to channel 2, and so on, with nodes subscribed to channel 4 changing to channel 1 (thus, the total amount of subscribers to each channel remains unchanged). We impose that nodes reset the counters associated to the items of the newly subscribed channel. This is reasonable, as it models the new personal interest of nodes for the channel. The figure presents the Hit Rate (a) and the overhead (b) from the point in time when subscriptions have changed. The best configuration of SCH (with RT=5) reaches the maximum HitRate before the best configuration of RH (with RT=75). Similarly to static scenarios, this is achieved with much lower overhead (which, in case of RH, keeps growing even after the HitRate stabilises). Reducing RT values limits the overhead of both SCH and RH, but the penalty in terms of reduced HitRate is much higher for RH.

Figure 5 presents the comparison between SCH and RH when a new channel, with its corresponding new items, is introduced in the network at time 10,000 sec. In this case, within all the communities, a majority of the nodes (15 over 25) subscribes to the new channel. Recognition levels of older items remain unchanged. The figure presents the Hit Rate of the nodes that changed their subscription and the overhead starting from $t = 10,000$. In this case, the advantage of SCH is even more visible. Specifically, SCH achieves a higher hit rate even with a very small RT value (RT=3). To get close to the SCH performance, we need to increase the RH value of RT at least to 30. But this results in a drastically higher overhead in terms of messages exchanged. Not only SCH achieves the same HitRate with far lower overhead, but the HitRate always increases faster (i.e., SCH achieves higher HitRate at any point in time). This is again due to the lower value of RT that SCH can use by still achieving 100% HitRate. With a lower RT value, old data items need not to be exchanged for long time, and therefore new data items do not have to compete with them for space in the nodes' caches. This manifests both with higher HitRates and lower overheads at any point in time.

Figure 6 compares SCH and RH when additional data items are created for existing channels. Specifically, we have doubled the number of data items for each channel at time 10,000 sec. After the injection of new items, SCH starts increasing the Hit Rate, nearly reaching 100% at the end of the simulation. Note that in this case the HitRate with RH increases much slower than before, and does not even match SCH also with a very high RT. Again, this is due to the fact that in RH new items have to compete with older data items that still have to reach the subscribed nodes, while this is not the case in SCH. In addition, as already observed in all the other experiments, RH requires significant higher overhead than SCH.

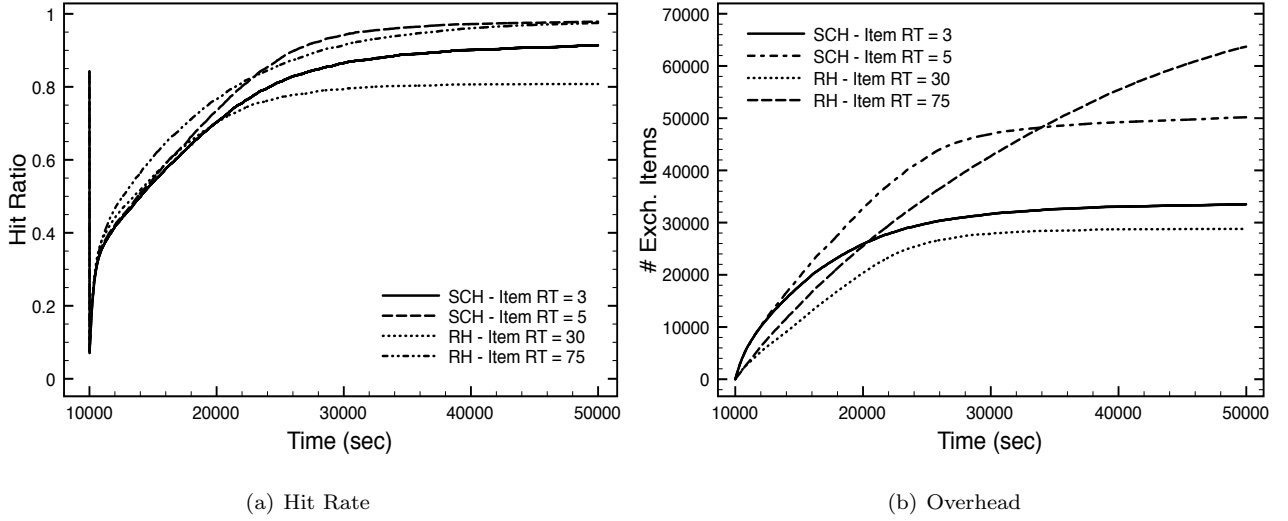

(a) Hit Rate

(b) Overhead

Figure 4: ZT Scenario - Hit Rate (a) and Overhead (b) with a dynamic change of subscriptions

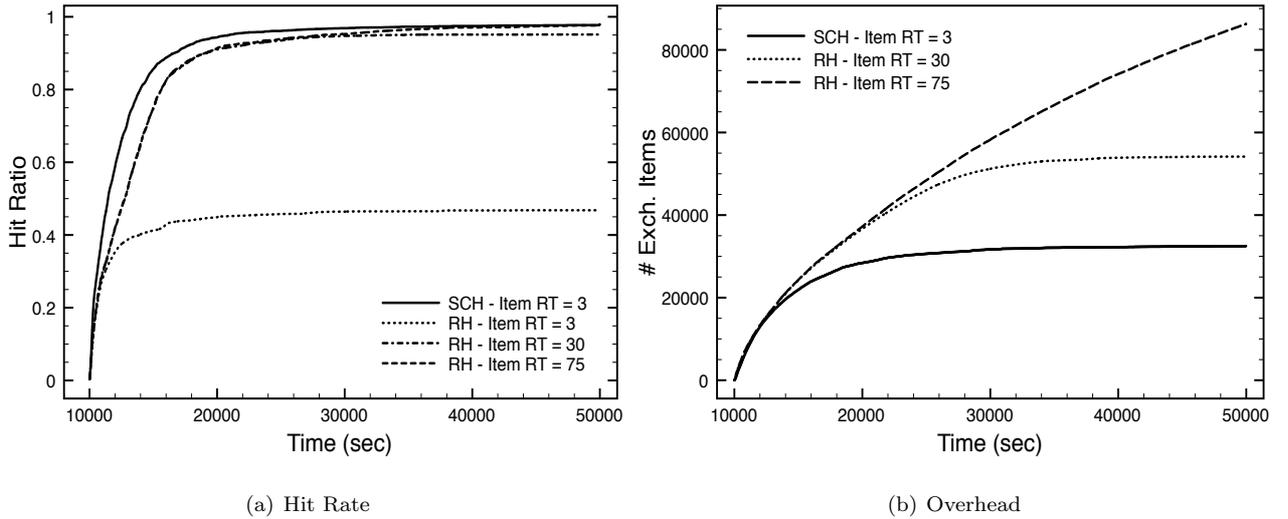

(a) Hit Rate

(b) Overhead

Figure 5: ZT Scenario - Hit Rate (a) and Overhead (b) with the dynamic creation of a new channel

We consider the latter dynamic scenario particularly realistic, as it represents conditions where new data items are generated for existing channels. Therefore, we investigate it further, by considering successive injections of new sets of data items at different points in time. In particular, we assume that new items are generated from $t = 10,000s$ and every 30,000 seconds afterwards. At each injection, 100 new items are created for every channel. They are placed uniformly at random in each community. Figure 7 shows the results of this experiment. At each injection, the Hit Rate temporarily drops, since new items suddenly appear. SCH is able to quickly react to the new situation and reaches a high Hit Rate after each data injection. Most importantly, the HitRate always increases from injection to injection, showing that SCH is able to bring all (old and new) data items to interested nodes. On the other hand, the HitRate of RH *degrades* at each injection. Because of the higher RTs that are needed to guarantee reasonable circulation of data items, old and new data items constantly compete against each other, and ultimately they are not delivered to interested nodes. In addition, the overhead is much higher than in SCH, and the difference constantly grows.

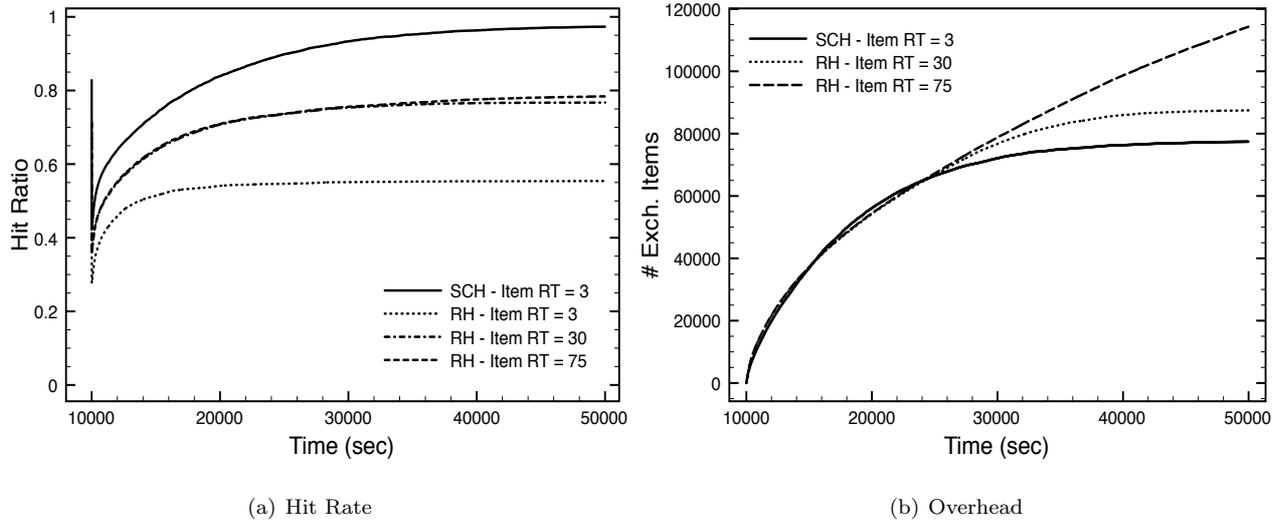

(a) Hit Rate

(b) Overhead

Figure 6: ZT Scenario - Hit Rate (a) and Overhead (b) with the dynamic injection of new items

## 7. Sensitiveness analysis

Other than the RT threshold, whose impact is already analysed in Sections 5 and 6, the other key parameter of our policies is the cache size. Therefore, in this section we analyse the impact of very limited amounts of storage space, to study the effectiveness of SCH also under very challenging conditions with respect to this parameter. The second parameter that we analyse is external to the data dissemination algorithms, and is specifically the initial distribution of data items in the communities. Finally, we analyse the performance when data needs to be delivered within a maximum time constraint, i.e., when requests for content have a Time-To-Live (TTL).

### 7.1. Sensitiveness to the Opportunistic Cache dimension

In this section we analyse the behaviour of SCH compared with RH when the dimension of the opportunistic cache is reduced. Smaller OCs could impact the dissemination process mainly for two reasons. First, a reduced OC dimension could slow down the dissemination process, since fewer data items can be delivered and exchanged by each node. Second, since only a small number of data items can be carried by a node, there is an increased impact of the mechanisms used by each peer to select the data to fetch at each encounter. Figures 8 and 9 show the comparison between SCH and RH when the dimension of the OC is set to 2 and 5, respectively, in the ZT scenario. Remember that in the results presented so far, the OC size was set to 10 messages. The other parameters are set to the standard values of the static conditions of the ZT scenario. When the OC has only two slots, both SCH and RH have not reached a convergence at the end of the simulation. However, it is worth noting that SCH is able to reach a slightly higher Hit Ratio with almost the same overhead of the two settings of RH. When the OC size is set to 5 (Figure 9), we can observe a behaviour that is qualitatively similar (see Figures 2(a) and 2(b)) to the one obtained with a larger OC: a slightly faster convergence with respect to RH (with item threshold 75), while requiring a much smaller overhead.

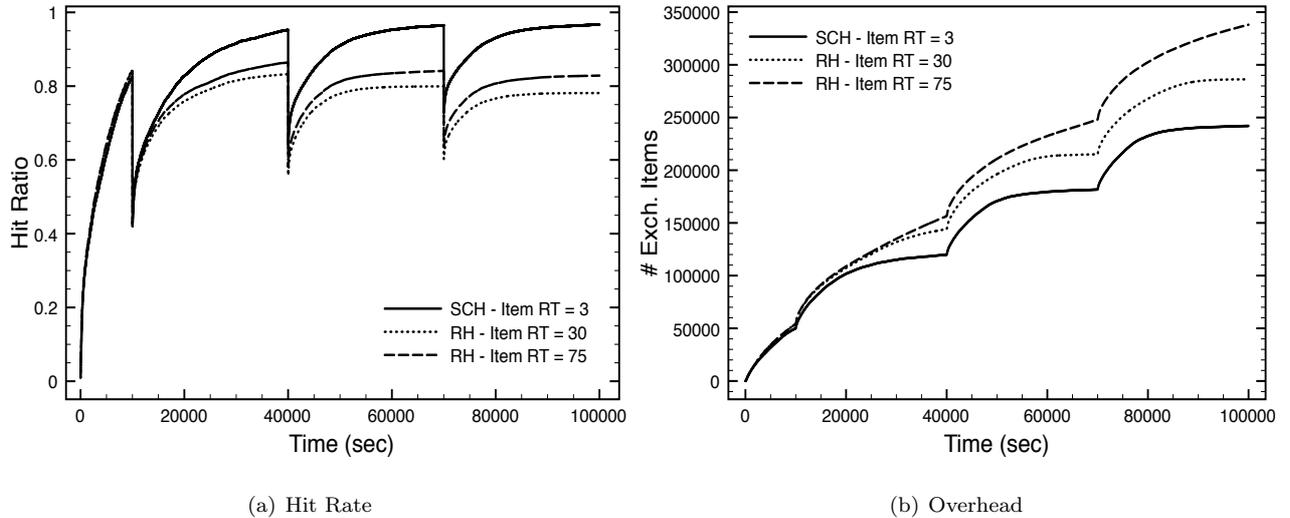

(a) Hit Rate

(b) Overhead

Figure 7: ZT Scenario - Hit Rate (a) and overhead (b) with a periodic injection of new items

These findings confirm the relevance of a social-based data selection scheme under resource constrained conditions, like a reduced space available for the dissemination task. In such cases, a sharp distinction among discovered data items is required to achieve an effective and efficient diffusion of the information.

### 7.2. Impact of different initial distributions of data items

While in all the previous results the data items generated at the beginning of the simulation were placed uniformly at random in the various groups of the system, in this subsection we we want to investigate how the initial data placement could affect the proposed data dissemination mechanisms. In the first configuration, we analyse the percentage of data of a channel generated in a group reflects the popularity of that channel in the group. Thus, the data of each channel is mainly generated in the group where the channel is the most popular. The other scenario presents the opposite situation, i.e. data is generated mainly in the group where its corresponding channel is the least popular one. Clearly, these configurations allow us to analyse the two extreme cases with respect to the analysed parameter. Figure 10 presents the results related to the first scenario, while Figure 11 shows the results of the second scenario.

The main difference between the two scenarios is that all the data dissemination schemes experience a faster intial growth of the Hit Ratio in the first scenario. This is due to the fact that peers of the most subscribed channels in each group can find the majority of their requested data directly inside their own groups. The Hit Ratio convergence time for SCH and RH is almost the same in both the scenarios. However, RH with an item recognition threshold of 30 it is not able to achieve 100% Hit Ratio in the second scenario.

In all the scenarios, SCH requires a noticeable lower overhead than RH. Moreover, in the second scenario (Figure 11) SCH requires a little lower overhead than in the first scenario (Figure 10 b) to obtain almost the same Hit Ratio performance. In this case, the majority of the subscirbers of a channel are placed in other groups with respect to those where the majority of the channel's data can be found. Therefore, the social-based data dissemination system in SCH plays an increasingly relevant role, since the ability of travellers to understand the needs of their fellow community members helps them to make a better selection of the items to fetch upon during contacts with other peers.

### 7.3. Performance with a limited TTL for data items

In next set of experiments, we want to understand how the system reacts when the validity of data items expires before the end of the simulation. This situation resembles scenarios where information is relevant in given time windows only, so that it should be timely delivered or discarded otherwise. In order to simulate this kind of scenarios, in the following we assume that each single data item is assigned a Time-To-Live

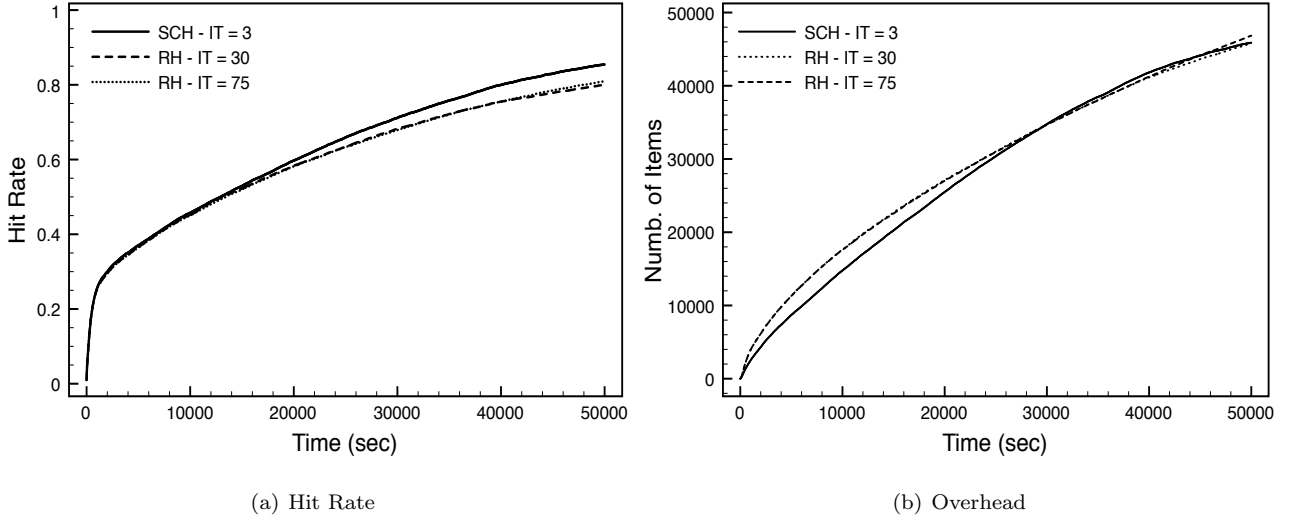

(a) Hit Rate

(b) Overhead

Figure 8: ZT Scenario - Hit Rate (a) and overhead (b) with an OC size = 2

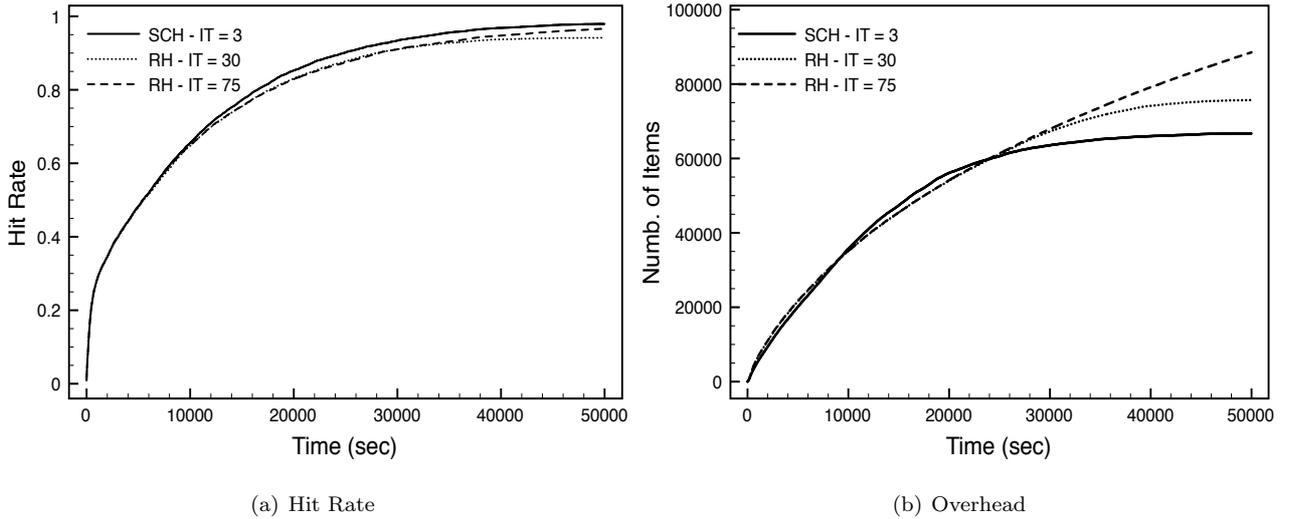

(a) Hit Rate

(b) Overhead

Figure 9: ZT Scenario - Hit Rate (a) and overhead (b) with with a OC size = 5

(TTL). In case the TTL expires, the item is removed from all the caches of the system. Thus, subscribed nodes that have not yet received this item do not have chances to retrieve it anymore. TTLs are assigned to data items according to a Gaussian distribution, with $\mu = T$ and $\sigma = 1500s$, where the values of $T$ are indicated in the following for each configuration. In these experiments, we do not expect the system to reach 100% of Hit Ratio, since many items are deleted before they can reach all their subscribers.

Figures 12, 13 and 14 present results obtained with mean TTLs (i.e., $T$) equal to $7500s$, $10000s$, and $15000s$, respectively. All the figures present qualitatively similar results. SCH and RH have no relevant differences in terms of Hit Ratio. This performance figure goes from 80%, with the shortest average TTL, to nearly 90% with the longest average TTL. As highlighted in almost all the previous experiments, SCH is able to achieve this Hit Ratio with a lower overhead than than RH in the corresponding competing settings.

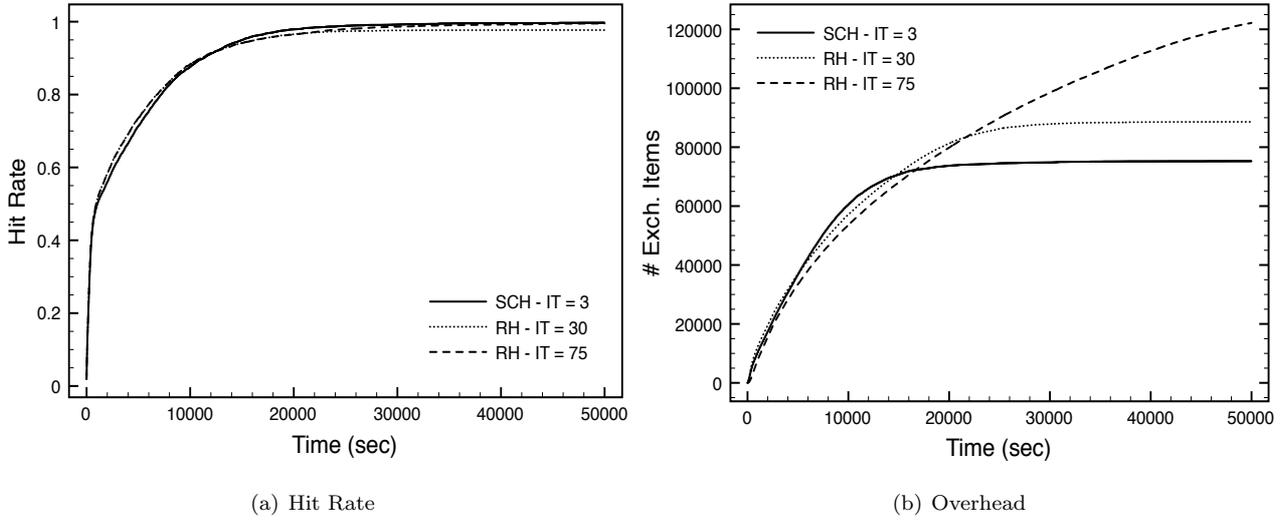

(a) Hit Rate

(b) Overhead

Figure 10: ZT Scenario - Hit Rate (a) and overhead (b) when data is initially placed according to the channel popularities

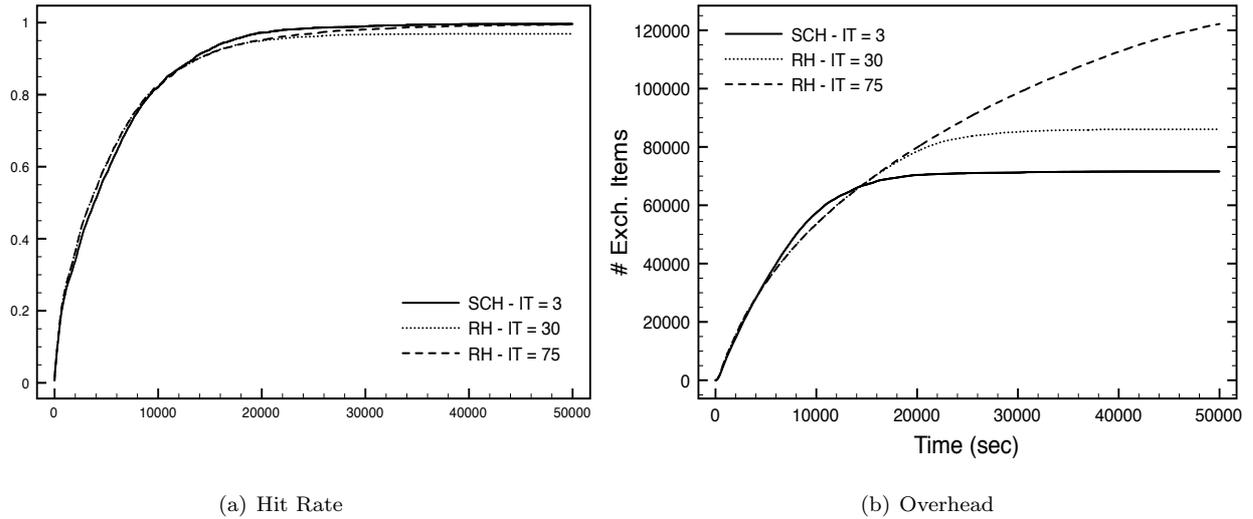

(a) Hit Rate

(b) Overhead

Figure 11: ZT Scenario - Hit Rate (a) and overhead (b) when data is initially placed in a proportion inversely correlated to the channel popularities

## 8. Impact of nodes collaboration and selfishness

In all the previous settings, we have assumed that nodes do not limit their participation in the cooperative data dissemination process. The next set of results shows the performance of the system in scenarios where peers adopt a partially selfish behaviour by limiting their engagement in the dissemination of information. A possible example of this kind of scenarios could be the necessity for a node to spare its own resources, like battery or bandwith. Therefore, in all the following experiments, peers do not accept to exchange information and data items with all the other nodes they meet. We present results exploiting different strategies to select the other nodes to interact with.

In the first of such strategies, each node has a uniform probability exchange contents upon meeting. Figures 15, 16 and 17 show the results when the *joint* exchange probability $p$ is 0.9, 0.5 and 0.33, respectively.

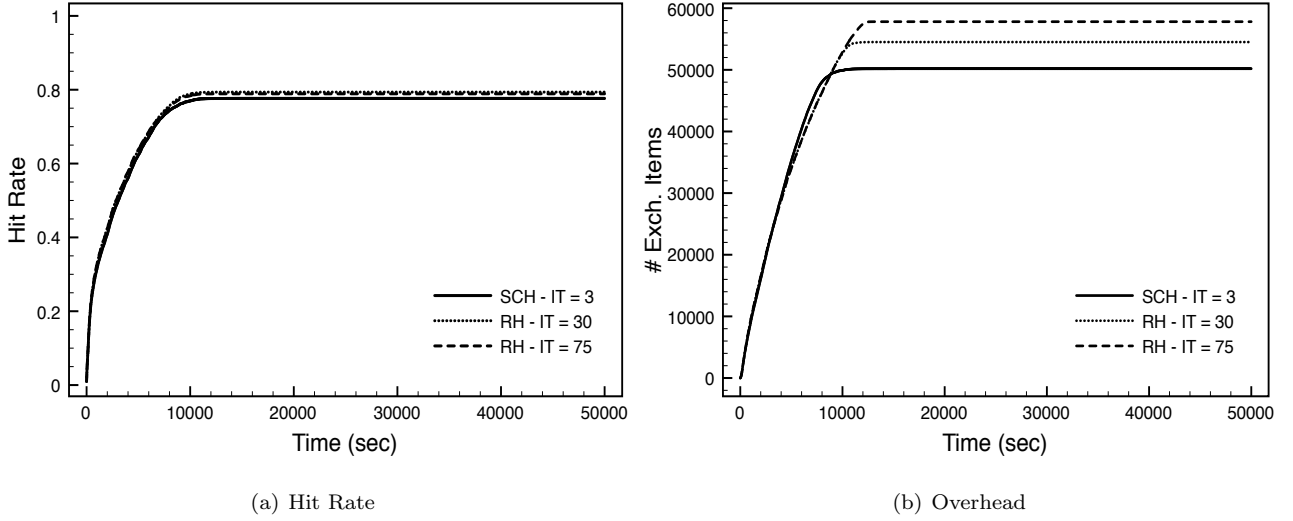

(a) Hit Rate

(b) Overhead

Figure 12: ZT Scenario - Hit Rate (a) and overhead (b) with a mean item TTL = 7500 sec.

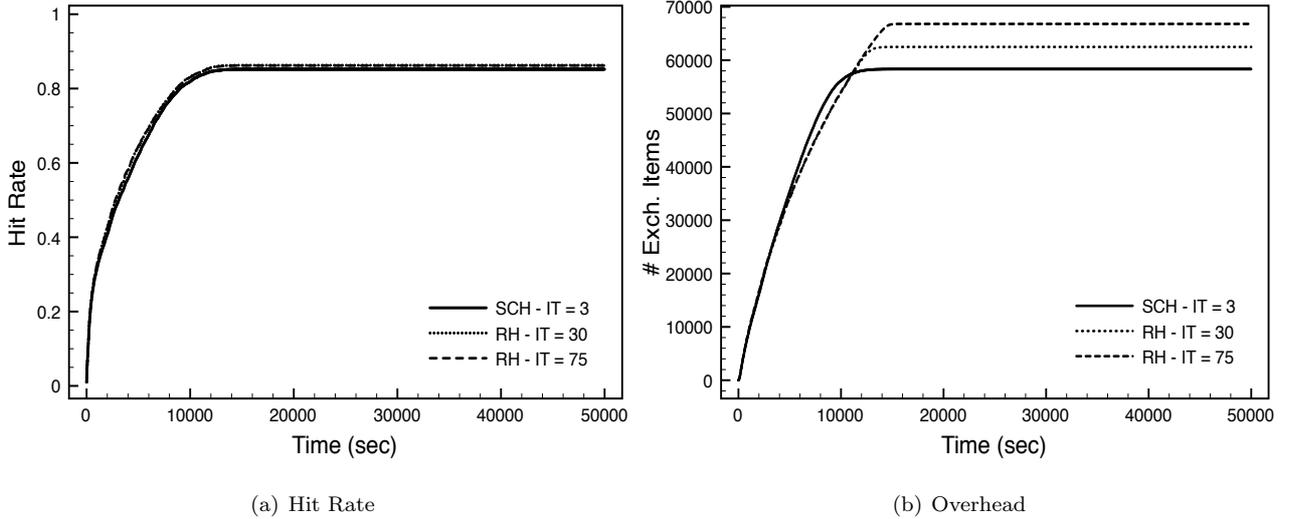

(a) Hit Rate

(b) Overhead

Figure 13: ZT Scenario - Hit Rate (a) and overhead (b) with a mean item TTL = 10000 sec.

Since we assume that all nodes behave the same, the probability for a single peer to be available for an exchange is $\sqrt{p}$. When $p = 0.9$ or $p = 0.5$, SCH obtains practically the same Hit Ratio performance of the competing settings of RH, with lower overheads. However, when $p = 0.33$, RH is not able anymore to reach a 100% Hit Ratio, while SCH maintains the ability to deliver all the data items to their subscribers. Furthermore, SCH continues to require much less overhead than RH.

This is a remarkable result. In fact, SCH achieves the maximum Hit Rate even when just one exchage over three effectively takes place, i.e. with $p = 0.33$. This result is also reached with a lower overhead than with the other values of $p$, as shown in Figure 18. Therefore, it is possible to take advantage of SCH for an effective and efficient dissemination of data even when the nodes exploit only a relatively small fraction of their encounters to effectively make exchanges. This fact opens the way for peers to balance the usage of their resources, while still giving a fruitful contribution to the collective data dissemination process.

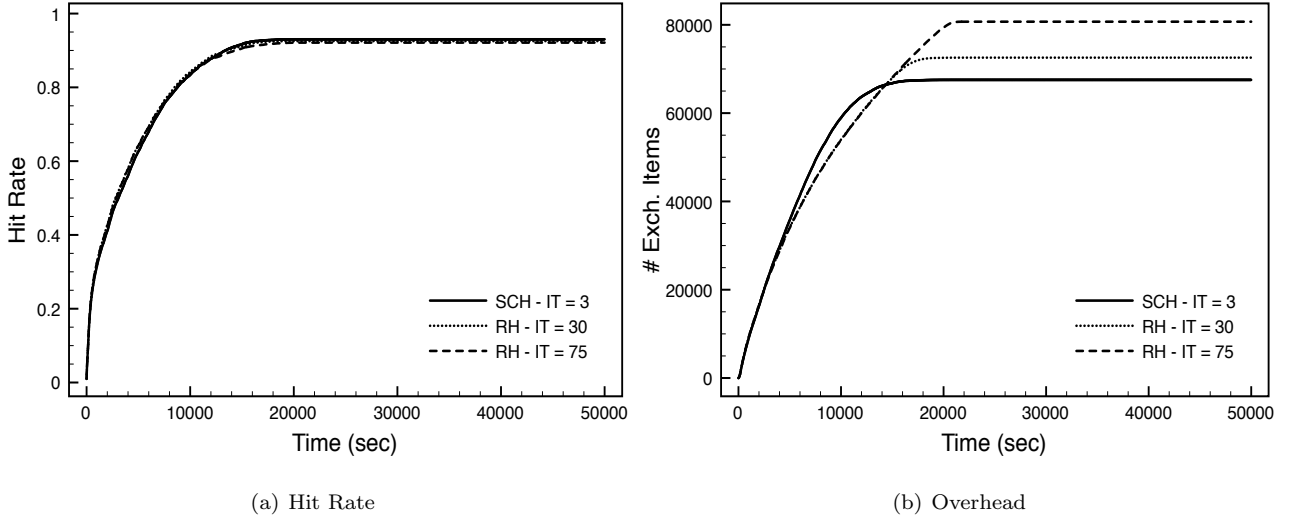

(a) Hit Rate

(b) Overhead

Figure 14: ZT Scenario - Hit Rate (a) and overhead (b) with a mean item TTL = 15000 sec.

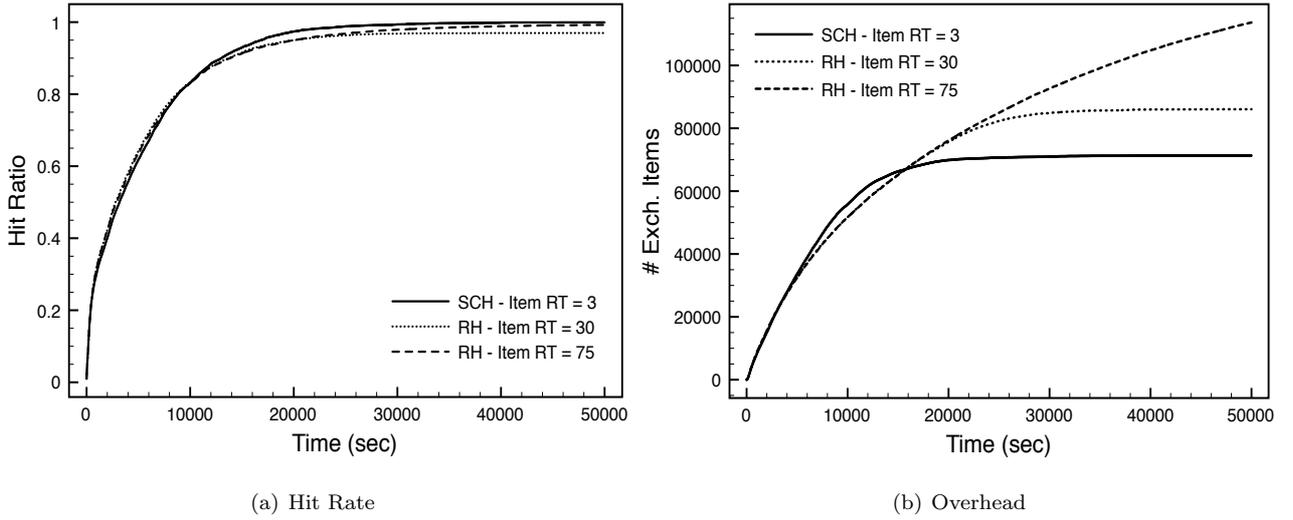

(a) Hit Rate

(b) Overhead

Figure 15: ZT Scenario - Hit Rate (a) and overhead (b) with a joint exchange probability $p = 0.9$

In the previous scenario, nodes use the same probability for all the other peers in order to decide whether they want to exchange data or not. However, SCH exploits a social strength detection mechanism. Leveraging on its knowledge about the strength of its social ties, a node could decide to assign different exchange probabilties to the members of its distinct social communities. Specifically, in the next experiments each node decides to be available for an exchange upon meeting with a device belonging to social group $i$ with a probability $p_i$. Specifically, we assume that each node ranks social communities (groups) based on the willingness of interacting and helping their members. The probability $p_i$ is defined as $p_i = 2^{-i}p_0$, where $p_0$ is the probability assigned to the group with which the node wants to interact most. Therefore, the members of the second preferred social group are accepted with half the proability of the first group, and so on. In the next experiments, for each node, the most preferred group is set either to the group that the nodes encounters most frequently, or less frequently, respectively. Our goal is to investigate whether is more efficient to make more exchanges with nodes that are met frequently (i.e., the most socially connected

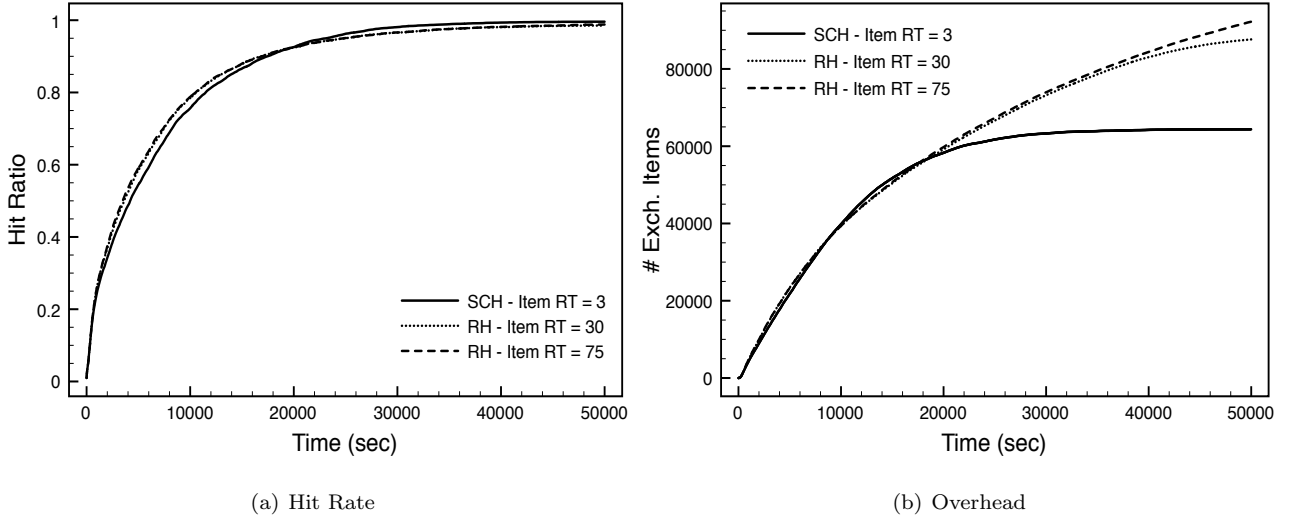

(a) Hit Rate

(b) Overhead

Figure 16: ZT Scenario - Hit Rate (a) and overhead (b) with a joint exchange probability $p = 0.5$

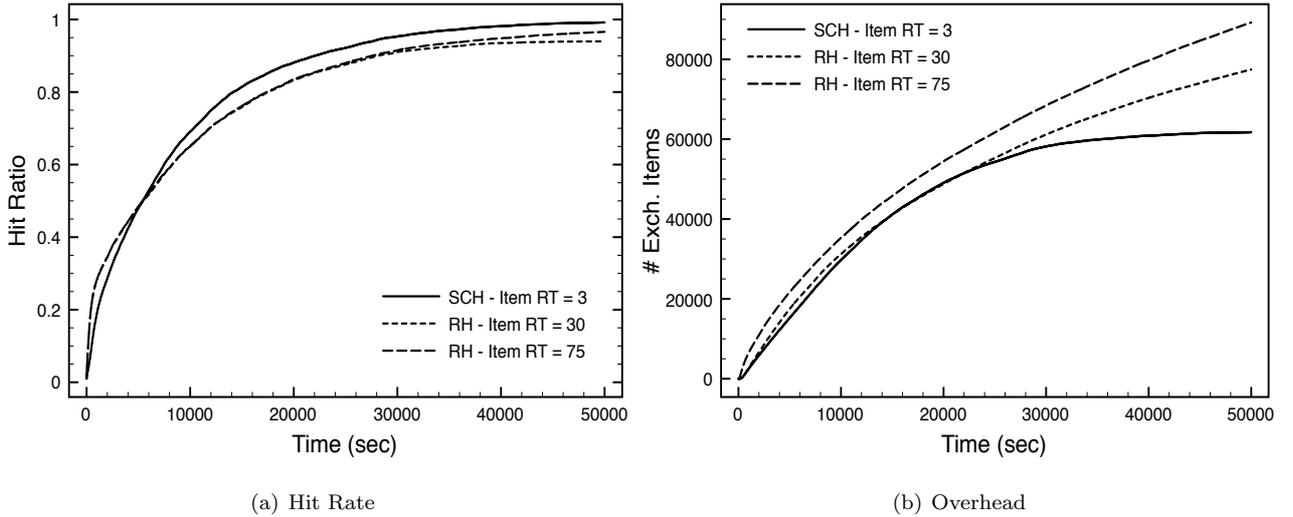

(a) Hit Rate

(b) Overhead

Figure 17: ZT Scenario - Hit Rate (a) and overhead (b) with a joint exchange probability $p = 0.33$

ones) or with nodes that are met only rarely (i.e., the ones with the lowest social connection).

Figure 19 shows the Hit Rate and the overhead of SCH when $p_0$ is assigned to the most socially connected group (social direct in the figure) or to the lowest socially connect group (social inverse in the figure). We also use two values for $p_0$, i.e. 1 and 0.75. When $p_0 = 1$, the data dissemination process is faster than when $p_0 = 0.75$ for both the social direct and the social inverse strategies. However, all the strategies lead to a final convergence of 100% Hit Rate. It is interesting to note, however, that the social inverse strategy always requires a final lower overhead than the social direct one. This fact can be explained in the following way. Less socially connected nodes are met less frequently. Therefore, it makes more sense to try to make more exchanges with them, since there are fewer events to do that. On the other hand, most socially connected peers have more possibilities to exchange data, since they are seen more frequently. Thus, even a lower exchange probabilities still gives them chances to pass and receive data items and information. In addition, it is known that social peers met not frequently act as "social bridges" between distant communities [46], and thus are more "precious" contacts for information diffusion than peers met frequently. With respect to the latter, as communities are typically tightly knit and most members encounter one another frequently, even dropping a large fraction of contact events does not leed to severe performance degradation.

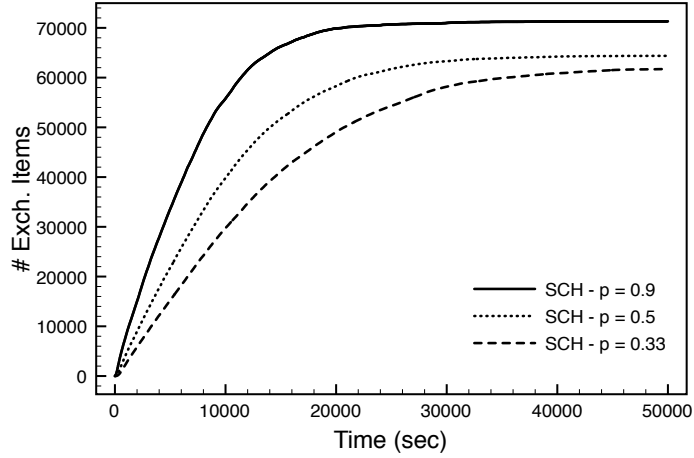

Figure 18: ZT Scenario - SCH overhead with different exchange probabilities

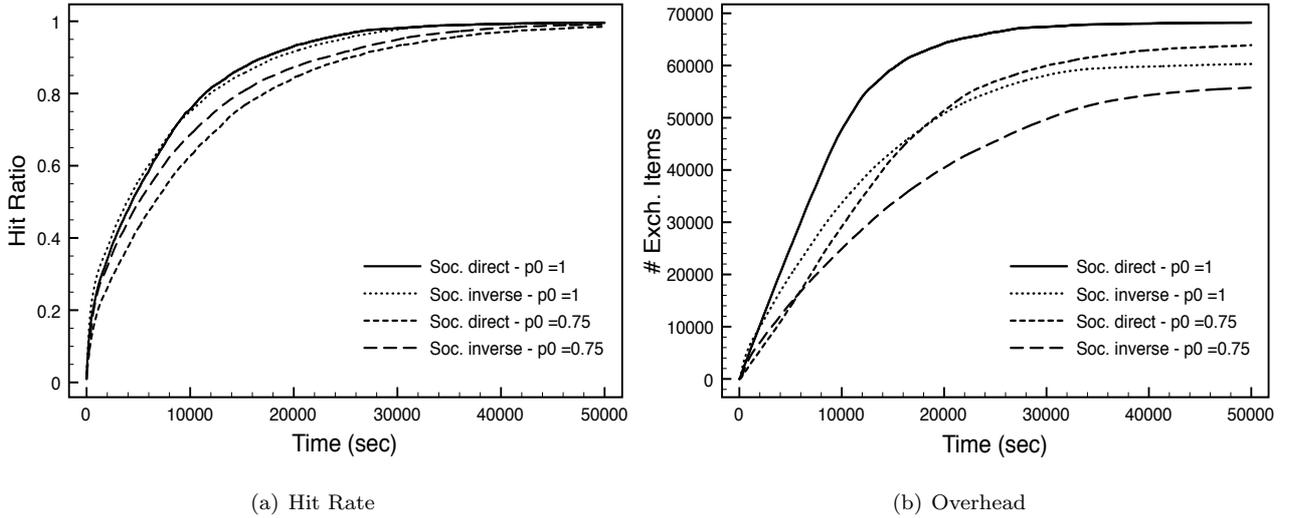

(a) Hit Rate

(b) Overhead

Figure 19: ZT Scenario - Hit Rate (a) and overhead (b) with different social-based exchange probabilities

In order to have a deeper uderstanding of the social-based exchange probability strategies, in the next experiments we report the result obtained from the comparison of these strategies with the ones based on uniform exchange probabilities. We also compare SCH to RH settings, in which case we cannot use any social information to select the peers to interact with, and thus assume a uniform random policy. In order to make social and uniform exchange proabilities comparable, the uniform exchange probability $p$ is determined as:

$$\sqrt{p} = \frac{\sum_{k=1}^{|G|} \bar{n}_k p_k}{\overline{N}} \qquad (2)$$

where $|G|$ is the number of social groups, $\bar{n}_k$ is the average number (over all the nodes in the system) of the members of the social group $k$, and $\overline{N} = \sum_{k=1}^{|G|} \bar{n}_k$ is the average number of social connections. Thus, $\sqrt{p}$ is the expected interaction probability of the SCH policy between any tagged node and all the other nodes it may encounter.

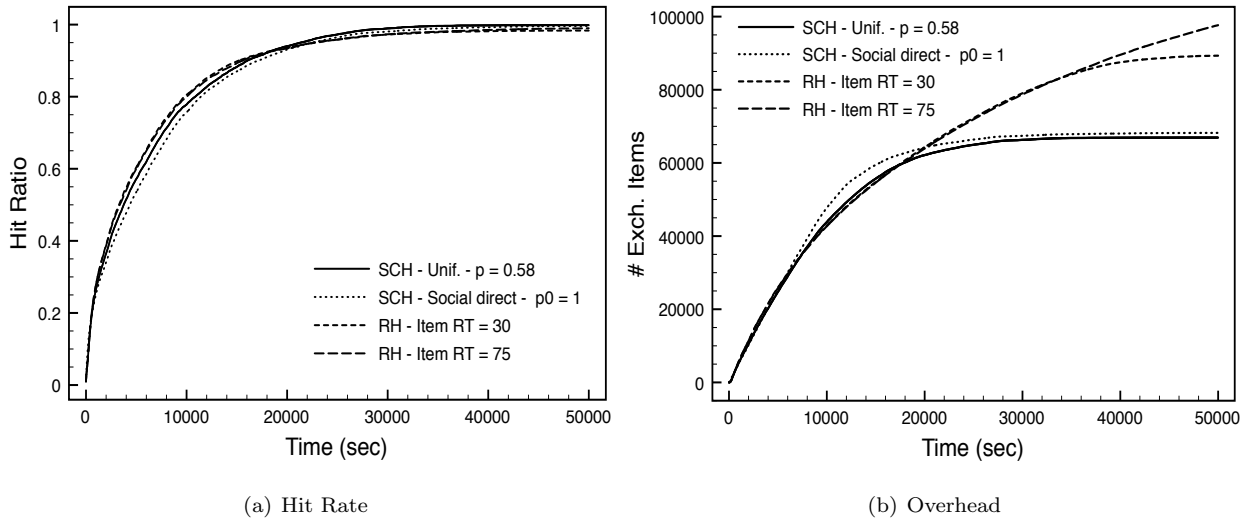

(a) Hit Rate

(b) Overhead

Figure 20: ZT Scenario - Hit Rate (a) and overhead (b) of a comparison of uniform and direct social-based ($p_0 = 1$) exchange probabilities

Figures 20 and 21 present the result of the comparison of the social direct strategy against the uniform ones. The value of $p$ is approximately 0.58 in Figure 20, and approximately 0.33 in Figure 21. The performance of the SCH social direct strategy with $p_0 = 1$ is essentially identical the SCH uniform one in all aspects, and both outperform the RH settings in terms of overhead. However, with $p_0 = 0.75$, the SCH uniform strategy tend to increase the Hit Rate at a faster pace than the social direct SCH scheme. It also requires a slightly lower overhead than the social direct SCH strategy. To explain this, consider that with these settings the probability to exchange data items with the least frequently encountered nodes is almost twice as high than in the social direct approach, while the probability to exchange data items with most frequently encountered nodes is approximately reduced by half. For the reasons highlighted in the previous experiments, increasing the probability for rarely met nodes is beneficial, while reducing the probability for frequently met nodes has little effect on the overall performance.

Figures 22 and 23 show the result of the comparison of the SCH social inverse strategy, when using $p_0 = 1$, and $p_0 = 0.75$, respectively. Differently from the previous case, these results show a striking difference between the social inverse strategy and all the other approaches. Specifically, the uniform exchange proabilites are approximately $p = 0.21$, and approximately $p = 0.12$, respectively. These values of $p$ are so low that all the uniform exchange-based schemes are not able to reach neither 100% Hit Rate, nor a state of convergence at the end of the simulation. On the other hand, SCH with the social inverse approach always achieves the maximum Hit Rate, with a considerable large difference with the other solutions when $p_0 = 0.75$. The overhead is larger than that of the other approaches just because they still need to exchange more items to reach the same Hit Rate of SCH social inverse.

From all these results, we can conclude that, when nodes behave selfishly to save resources, it is more efficient if they prefer to exploit more rare contacts with socially distant nodes, as opposed to prefer more frequent contacts with socially close nodes. With respect to the opposite approach of preferring socially close nodes, this achieves comparable hit rates with reduced overheads. With respect to uniform policies, this achieves higher Hit Rate at the cost of some additional overhead.

## 9. Conclusion

In this paper we have faced the problem of data dissemination for Future Internet systems using a self-organising approach based on D2D technologies. Starting from the general idea that data-centric services running on personal mobile devices need to incorporate algorithms that closely match the behaviuor of their human users in the physical world, we have proposed a novel data dissemination scheme for Opportunistic Networks built on cognitive heuristics, i.e. models of the human cognitive processes defined in the cognitive psychology literature. The proposed solution is built on a *social* cognitive heuristic, the Social Circle Heuristic (SCH). According to the SCH definition, the human brain evaluates relevance of information based not only on its own judgement, but also on the judgement of its social contacts, taken in order of social proximity. The proposed data dissemination scheme uses the very same algorithmic description of SCH. The proposed scheme is compared against alternative non-social cognitive solutions, that have been shown to be very effective in recent work. Results clearly show the advantage brought by using social cognitive heuristics. Already in static scenarios (when content and interests of users do not change) social cognitive schemes are as effective as non-cognitive schemes, but generate significantly lower overhead. This advantage is further confirmed by varying the various parameters that characterise the system. Moreover, in more dynamic (and realistic) scenarios where either new content is generated or users' interests change over time, social cognitive schemes outperform non-social scheme to an even greater extent. Specifically, we have found that when new content is periodically generated, non-social schemes enter a saturation condition where new and old data items compete for being disseminated, and this results in *higher* overhead with *lower and decreasing* effectiveness in data dissemination, with respect to social cognitive schemes. An even greater impact is shown by the social-based data dissemination scheme in contexts where nodes do not fully collaborate to the dissemination process, e.g. due to resource-saving or selfishness. In particular, a version of SCH that gives preference to contacts with socially distant nodes to disseminate data has proven to achieve much better results than non-social based schemes.

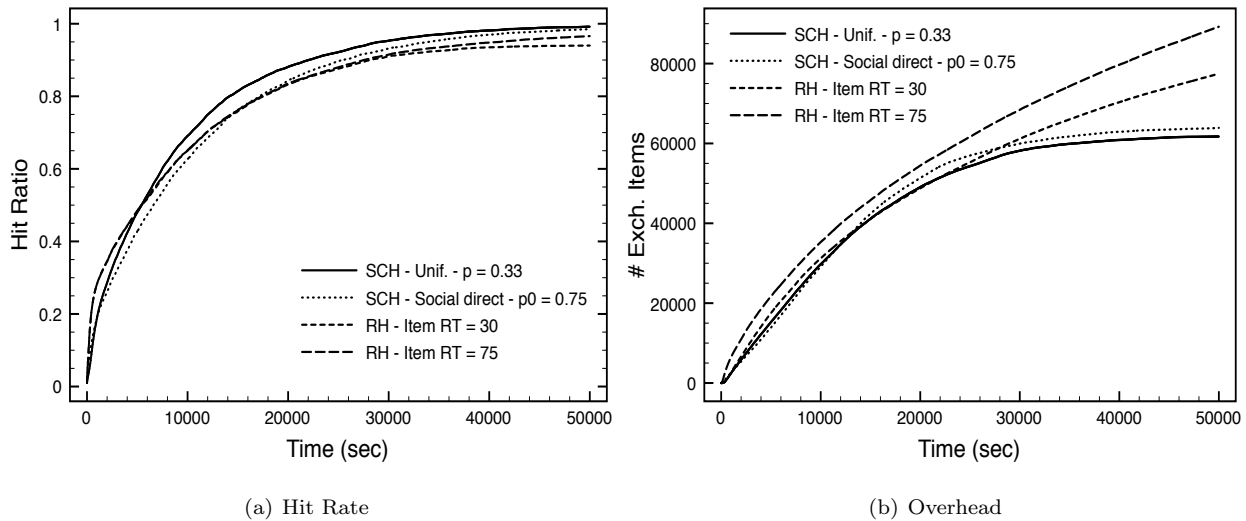

(a) Hit Rate  (b) Overhead

Figure 21: ZT Scenario - Hit Rate (a) and overhead (b) of a comparison of uniform and direct social-based ($p_0 = 0.75$) exchange probabilities

## Acknowledgments

This work is partly funded by the EC under the H2020 REPLICATE (691735), SoBigData (654024) and AUTOWARE (723909) projects.## References

[1] S. Trifunovic, S. T. Kouyoumdjieva, B. Distl, L. Pajevic, G. Karlsson, B. Plattner, A decade of research in opportunistic networks: Challenges, relevance, and future directions, IEEE Communications Magazine 55 (1) (2017) 168–173.25

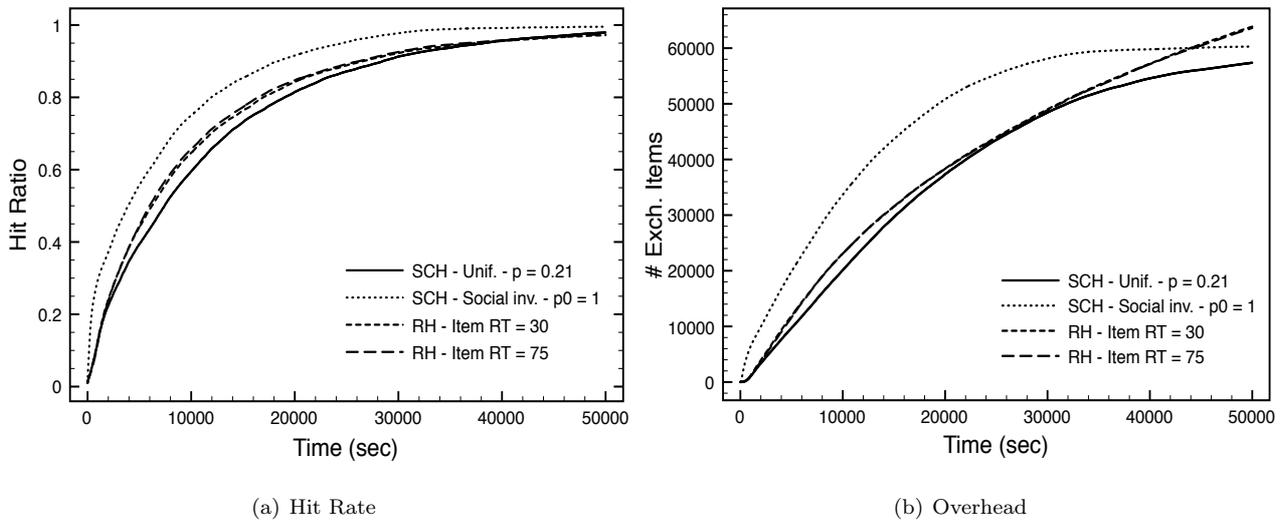

(a) Hit Rate

(b) Overhead

Figure 22: ZT Scenario - Hit Rate (a) and overhead (b) of a comparison of uniform and inverse social-based ($p_0 = 1$) exchange probabilities

[2] M. Conti, C. Boldrini, S. S. Kanhere, E. Mingozzi, E. Pagani, P. M. Ruiz, M. Younis, From manet to people-centric networking: milestones and open research challenges, Computer Communications 71 (2015) 1–21.
[3] I. F. Akyildiz, W. Su, Y. Sankarasubramaniam, E. Cayirci, Wireless sensor networks: a survey, Computer networks 38 (4) (2002) 393–422.
[4] M. Amjad, M. Sharif, M. K. Afzal, S. W. Kim, Tinyos-new trends, comparative views, and supported sensing applications: A review, IEEE Sensors Journal 16 (9) (2016) 2865–2889.
[5] G. Gigerenzer, D. G. Goldstein, Reasoning the fast and frugal way: models of bounded rationality., Psychological review 103 (4) (1996) 650–69.
[6] G. Gigerenzer, P. M. Todd, Simple Heuristics that Make Us Smart, Oxford University Press, New York, 1999.
[7] M. Conti, S. K. Das, C. Bisdikian, M. Kumar, L. M. Ni, A. Passarella, G. Roussos, G. Tröster, G. Tsudik, F. Zambonelli, Looking ahead in pervasive computing: Challenges and opportunities in the era of cyber–physical convergence, Pervasive and Mobile Computing 8 (1) (2012) 2–21.
[8] M. Conti, M. Mordacchini, A. Passarella, Design and performance evaluation of data dissemination systems for opportunistic networks based on cognitive heuristics, ACM Trans. Auton. Adapt. Syst. 8 (3) (2013) 12:1–12:32. doi:10.1145/2518017.2518018.
[9] M. Mordacchini, A. Passarella, M. Conti, S. M. Allen, M. J. Chorley, G. B. Colombo, V. Tanasescu, R. M. Whitaker, Croudsourcing through cognitive opportunistic networks, ACM Trans. Auton. Adapt. Syst. 10 (2) (2015) 13:1,13:30.
[10] M. Mordacchini, L. Valerio, M. Conti, A. Passarella, A cognitive-based solution for semantic knowledge and content dissemination in opportunistic networks, in: Proc. of AOC 2013, IEEE, 2013.
[11] M. Mordacchini, L. Valerio, M. Conti, A. Passarella, Design and evaluation of a cognitive approach for disseminating semantic knowledge and content in opportunistic networks, Computer Communications 81 (2016) 12–30.
[12] C. Boldrini, M. Conti, A. Passarella, Design and performance evaluation of ContentPlace, a social-aware data dissemination system for opportunistic networks, Comput. Netw. 54 (2010) 589–604.
[13] X. Zhuo, Q. Li, G. Cao, Y. Dai, B. Szymanski, T. Porta, Social-based cooperative caching in dtns: A contact duration aware approach, in: Proc. of IEEE MASS, 2011.
[14] E. Yoneki, P. Hui, S. Chan, J. Crowcroft, A socio-aware overlay for publish/subscribe communication in delay tolerant networks, in: Proceedings of the 10th ACM MSWiM Conf., 2007.
[15] T. Pachur, J. Rieskamp, R. Hertwig, The social circle heuristic: Fast and frugal decisions based on small samples, in: Proc. of the 26th Annual Conf. of the Cognitive Science Soc., Erlbaum, 2005, pp. 1077–1082.
[16] T. Pachur, L. J. Schooler, J. R. Stevens, We'll meet again: Revealing distributional and temporal patterns of social contact, PLoS ONE 9 (1) (2014) e86081.
[17] R. Dunbar, {The social brain hypothesis}, Evol. Anthrop. 6 (1998) 178–190.
[18] E. Cho, S. A. Myers, J. Leskovec, Friendship and mobility: User movement in location-based social networks, in: Proc. of ACM KDD, 2011.
[19] S. Scellato, A. Noulas, R. Lambiotte, C. Mascolo, Socio-spatial properties of online location-based social networks, in: Proc. of ICWSM, 2011.
[20] S. Scellato, A. Noulas, C. Mascolo, Exploiting place features in link prediction on location-based social networks, in: Proc. of ACM KDD, 2011.
[21] L. Valerio, M. Conti, E. Pagani, A. Passarella, Autonomic cognitive-based data dissemination in opportunistic networks,

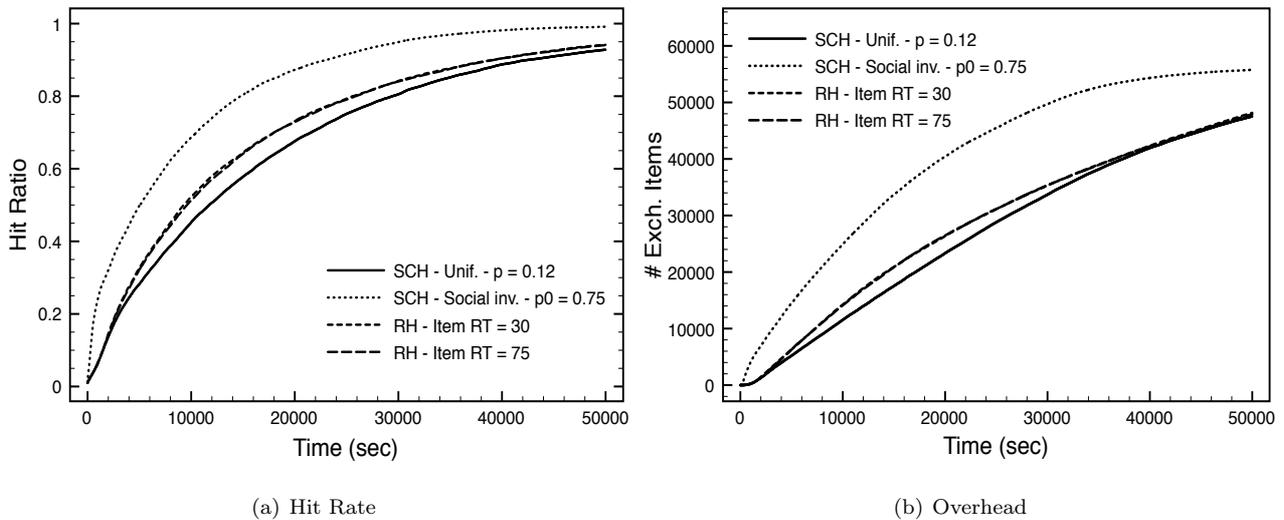

Figure 23: ZT Scenario - Hit Rate (a) and overhead (b) of a comparison of uniform and inverse social-based ($p_0 = 0.75$) exchange probabilities

in: Proc. of IEEE WOWMOM 2013, 2013, pp. 1–10.
[22] R. Bruno, M. Conti, M. Mordacchini, A. Passarella, An analytical model for content dissemination in opportunistic networks using cognitive heuristics, in: Proc. of MSWIM 2012, ACM, 2012, pp. 61–68.
[23] A. Bujari, A survey of opportunistic data gathering and dissemination techniques, in: Proc. of ICCCN 2012, IEEE, 2012, pp. 1–6.
[24] C. Boldrini, A. Passarella, Data dissemination in opportunistic networks, in: S. Basagni, M. Conti, S. Giordano, I. Stojmenovic (Eds.), Mobile Ad Hoc Networking: Cutting Edge Directions, Second Edition, John Wiley & Sons, Inc., 2013, pp. 453–490.
[25] V. Lenders, M. May, G. Karlsson, C. Wacha, Wireless ad hoc podcasting, SIGMOBILE Mob. Comput. Commun. Rev. 12 (2008) 65–67.
[26] H. Zhou, J. Wu, H. Zhao, S. Tang, C. Chen, J. Chen, Incentive-driven and freshness-aware content dissemination in selfish opportunistic mobile networks, IEEE Transactions on Parallel and Distributed Systems 26 (9) (2015) 2493–2505.
[27] R.-I. Ciobanu, R.-C. Marin, C. Dobre, V. Cristea, Interest-awareness in data dissemination for opportunistic networks, Ad Hoc Networks 25 (2015) 330–345.
[28] J. Reich, A. Chaintreau, The age of impatience: optimal replication schemes for opportunistic networks, in: Proc. of the 5th international conference on Emerging networking experiments and technologies, CoNEXT '09, ACM, New York, NY, USA, 2009, pp. 85–96.
[29] Y. Liu, H. Wu, Y. Xia, Y. Wang, F. Li, P. Yang, Optimal online data dissemination for resource constrained mobile opportunistic networks, IEEE Transactions on Vehicular Technology PP (99) (2016) 1–1. doi:10.1109/TVT.2016.2616034.
[30] I. H. Chuang, T. A. Lin, C. C. Lo, Y. H. Kuo, Time-sensitive data dissemination in opportunistic networks, in: 2014 IEEE Wireless Communications and Networking Conference (WCNC), 2014, pp. 2420–2425. doi:10.1109/WCNC.2014.6952728.
[31] M. Mordacchini, A. Passarella, M. Conti, Social cognitive heuristics for adaptive data dissemination in opportunistic networks, in: World of Wireless, Mobile and Multimedia Networks (WoWMoM), 2015 IEEE 16th International Symposium on a, IEEE, 2015, pp. 1–9.
[32] T. Pachur, R. Hertwig, J. Rieskamp, Intuitive judgements of social statistics: How exhaustive does sampling need to be?, J. of Exp. Psych. 49 (2013) 1059–1077.
[33] G. Gigerenzer, D. G. Goldstein, Models of ecological rationality: The recognition heuristic, Psychological Review 109 (1) (2002) 75–90.
[34] W.-X. Zhou, D. Sornette, R. A. Hill, R. I. Dunbar, Discrete hierarchical organization of social group sizes, Proc. of the Royal Society B: Biological Sciences 272 (1561) (2005) 439–444.
[35] A. Sutcliffe, R. Dunbar, J. Binder, H. Arrow, Relationships and the social brain: Integrating psychological and evolutionary perspectives, British journal of psychology 103 (2) (2012) 149–168.
[36] J. N. Marewski, W. Gaissmaier, L. J. Schooler, D. G. Goldstein, G. Gigerenzer, From recognition to decisions: Extending and testing recognition-based models for multialternative inference, Psychonomic Bulletin & Review 17 (3) (2010) 287–309.
[37] M. Mordacchini, A. Passarella, M. Conti, Community detection in opportunistic networks using memory-based cognitive heuristics, in: Proc. of the IEEE 3rd Int. Workshop PerMoby 2014, IEEE, 2014.
[38] T. Pachur, L. J. Schooler, J. R. Stevens, When Will We Meet Again? Regularities of Social Connectivity and Their Reflections in Memory and Decision Making, Oxford University Press, 2012. doi:10.1093/acprof:oso/9780195388435.003.0007.

[39] D. Read, Y. Grushka-Cockayne, The Similarity Heuristic, Journal of Behavioral Decision Making (24) (2011) 23–46. doi:10.1002/bdm.
[40] C. Boldrini, A. Passarella, Hcmm: Modelling spatial and temporal properties of human mobility driven by users' social relationships, Comput. Commun. 33 (2010) 1056–1074. doi:http://dx.doi.org/10.1016/j.comcom.2010.01.013.
[41] P. Pirozmand, G. Wu, B. Jedari, F. Xia, Human mobility in opportunistic networks: Characteristics, models and prediction methods, Journal of Network and Computer Applications 42 (2014) 45–58.
[42] X. Hu, T. H. Chu, V. C. Leung, E. C.-H. Ngai, P. Kruchten, H. C. Chan, A survey on mobile social networks: Applications, platforms, system architectures, and future research directions, IEEE Communications Surveys & Tutorials 17 (3) (2015) 1557–1581.
[43] D. Karamshuk, C. Boldrini, M. Conti, A. Passarella, Human mobility models for opportunistic networks, IEEE Communications Magazine 49 (12) (2011) 157–165.
[44] W. Navidi, T. Camp, Stationary distributions for the random waypoint mobility model, IEEE transactions on Mobile Computing 3 (1) (2004) 99–108.
[45] L. Sun, W. Wang, Y. Li, The impact of network size and mobility on information delivery in cognitive radio networks, IEEE Transactions on Mobile Computing 15 (1) (2016) 217–231.
[46] M. S. Granovetter, The strength of weak ties, American journal of sociology (1973) 1360–1380.